\documentclass[onecolumn,showpacs,11pt,nofootinbib]{revtex4-2}
\pdfoutput=1
\usepackage{axodraw}
\usepackage{graphicx}
\usepackage{dcolumn}
\usepackage{bm}
\usepackage{color}
\usepackage{eurosym}
\usepackage{multirow}
\usepackage{amsfonts}
\usepackage{amsmath}
\usepackage{hyperref}
\usepackage{amssymb}
\usepackage[english]{babel}
\usepackage{graphicx}
\usepackage{epsfig}
\usepackage{subfigure}
\usepackage{unicode}
\begin{document}
\newcommand{\hs}{\hspace*{0.2cm}}
\newcommand{\hsp}{\hspace*{0.5cm}}
\newcommand{\vs}{\vspace*{0.5cm}}
\newcommand{\be}{\begin{equation}}
\newcommand{\ee}{\end{equation}}
\newcommand{\bea}{\begin{eqnarray}}
\newcommand{\eea}{\end{eqnarray}}
\newcommand{\ben}{\begin{enumerate}}
\newcommand{\een}{\end{enumerate}}
\newcommand{\bde}{\begin{widetext}}
\newcommand{\ede}{\end{widetext}}
\newcommand{\nn}{\nonumber}
\newcommand{\crn}{\nonumber \\}
\newcommand{\Tr}{\mathrm{Tr}}
\newcommand{\noi}{\noindent}
\newcommand{\al}{\alpha}
\newcommand{\la}{\lambda}
\newcommand{\bet}{\beta}
\newcommand{\ga}{\gamma}
\newcommand{\va}{\varphi}
\newcommand{\om}{\omega}
\newcommand{\pa}{\partial}
\newcommand{\+}{\dagger}
\newcommand{\fr}{\frac}
\newcommand{\sq}{\sqrt}
\newcommand{\bc}{\begin{center}}
\newcommand{\ec}{\end{center}}
\newcommand{\Ga}{\Gamma}
\newcommand{\de}{\delta}
\newcommand{\De}{\Delta}
\newcommand{\ep}{\epsilon}
\newcommand{\varep}{\varepsilon}
\newcommand{\ka}{\kappa}
\newcommand{\La}{\Lambda}
\newcommand{\si}{\sigma}
\newcommand{\Si}{\Sigma}
\newcommand{\ta}{\tau}
\newcommand{\up}{\upsilon}
\newcommand{\Up}{\Upsilon}
\newcommand{\ze}{\zeta}
\newcommand{\ps}{\psi}
\newcommand{\Ps}{\Psi}
\newcommand{\ph}{\phi}
\newcommand{\vph}{\varphi}
\newcommand{\Ph}{\Phi}
\newcommand{\Om}{\Omega}
\newcommand{\Revised}[1]{{\color{red}#1}}
\newcommand{\Red}[1]{{\color{red}#1}}
\newcommand{\crb}[1]{{\color{blue}#1}}
\title{Multiscalar $B-L$ extension based on $S_4$ flavor symmetry\\ for neutrino mass and mixing}
\author{V. V. Vien$^{a,b}$}
\email{wvienk16@gmail.com}
\author{H. N. Long$^{c,d}$}
\email{hoangngoclong@tdtu.edu.vn}
\affiliation{$^{a}$Institute of Research and Development, Duy Tan University, 182
Nguyen Van Linh, Da Nang City, Vietnam\\
$^{b}$Department of Physics, Tay Nguyen University, 567 Le Duan, Buon Ma
Thuot, DakLak, Vietnam. \\
$^c$Theoretical Particle Physics and Cosmology Research Group,
Advanced Institute of Materials Science, Ton Duc Thang University, Ho Chi
Minh City, Vietnam\\
$^d$Faculty of Applied Sciences, Ton Duc Thang University, Ho Chi Minh City,
Vietnam}
\begin{abstract}
 A multiscalar and nonrenormalizable $B-L$ extension of the standard model (SM) with $S_4$ symmetry which successfully explains
the recent observed neutrino oscillation data is proposed. The tiny neutrino
masses and their hierarchies are generated via the type-I seesaw mechanism.
The model reproduces the recent experiments of neutrino mixing angles and  Dirac CP violating phase in which the
atmospheric angle $(\theta_{23})$ and the reactor angle $(\theta_{13})$ get the best-fit values while the solar angle $(\theta_{12})$
	 and Dirac CP violating phase ($\de $) belong to $3\, \si $ range of the best-fit value for normal hierarchy (NH). For inverted
	  hierarchy (IH), $\theta_{13}$ gets the best-fit value and $\theta_{23}$ together with $\de $ belongs to $1\, \si $ range
	   while $\theta_{12}$ belongs to $3\, \si $ range of the best-fit value. The effective neutrino masses are predicted
  to be $\langle m_{ee}\rangle=6.81 \,\, \mbox{meV}$ for NH and $\langle m_{ee}\rangle=48.48\,\, \mbox{meV}$
  for IH being in good agreement with the most recent experimental data.
\end{abstract}
\date{\today}
\maketitle

\section{\label{intro} Introduction}
The observed neutrino oscillation data, including the neutrino mass-squared differences, mixing angles as well as the Dirac CP phase
 given in Table \ref{tab1},  is one of the most appealing subjects of current Particle Physics.
  This pattern provides inspiration for constructing
models with additional scalars and symmetries and makes it interesting to extend the SM.
\begin{table}[ht]
\caption{\label{tab1} Observed neutrino oscillation data taken from Ref. \cite{Esteban2020}. Here, $\De  m^2_{3l}\equiv \De  m^2_{31} >0$
for NH and $\De  m^2_{3l}\equiv \De  m^2_{32} < 0$ for IH.}\label{1}
\vspace{0.15cm}
\begin{tabular}{|c|c|c|c|c|c|c|c|}
\hline
\multirow{2}{2.3cm}{\hfill Parameters   \hfill } & NH &
$\mathrm{IH}$ \\
\cline{2-3}   & $\mathrm{bfp}\pm 1\si $\hspace{0.75cm}  $3\si $ range&
$\mathrm{bfp}\pm 1\si $\hspace{0.5cm} $3\si $ range \\  \hline
   $\sin^2\theta_{23}$& $0.573_{-0.020}^{+0.016}$ \hspace{0.75cm} $0.415 \to 0.616$&
  $0.575_{-0.019}^{+0.016}$ \hspace{0.75cm} $0.419 \to 0.617$\\
  $\theta_{23}/^\circ$& $49.2_{-1.2}^{+0.9}$ \hspace{0.75cm} $40.1 \to 51.7$    &
  $49.3_{-1.1}^{+0.9}$ \hspace{0.75cm} $40.3 \to
  51.8$\\
      $\sin^2\theta_{12}$& $0.304_{-0.012}^{+0.012}$ \hspace{0.75cm} $0.269 \to 0.343$ & $0.304_{-0.012}^{+0.013}$
       \hspace{0.75cm} $0.269 \to 0.343$ \\
      $\theta_{12}/^\circ$      & $33.44_{-0.74}^{+0.77}$ \hspace{0.75cm} $31.27 \to 35.86$ &
       $33.45_{-0.75}^{+0.78}$\hspace{0.5cm} $31.27 \to 35.87$      \\
       $\sin^2\theta_{13}$   & $0.02219_{-0.00063}^{+0.00062}$ \hspace{0.75cm} $0.02032 \to 0.02410$ &
        $0.02238_{-0.00062}^{+0.00063}$ \hspace{0.75cm} $0.02052 \to 0.02428$      \\
       $\theta_{13}/^\circ$  & $8.57_{-0.12}^{+0.12}$ \hspace{0.75cm} $8.20 \to 8.93$  & $8.60_{-0.12}^{+0.12}$
        \hspace{0.75cm} $8.24 \to 8.96$      \\
      $\de _{CP}(^\circ)$& $197_{-24}^{+27}$ \hspace{0.75cm} $120 \to 369$      & $282_{-30}^{+26}$
      \hspace{0.5cm} $193 \to 352$ \\
      $\fr{\De  m^2_{21}}{10^{-5}\, \mathrm{eV}^2}$& $7.42_{-0.20}^{+0.21}$ \hspace{0.75cm} $6.82 \to 8.04$
      & $7.42_{-0.20}^{+0.21}$ \hspace{0.75cm} $6.82 \to 8.04$
      \\
       $\fr{\De  m^2_{3 l}}{10^{-3}\, \mathrm{eV}^2}$& $+2.517_{-0.028}^{+0.026}$
       \hspace{0.75cm} $+2.435 \to +2.598$
      & $-2.498_{-0.028}^{+0.028}$ \hspace{0.75cm} $-2.581 \to -2.414$      \\     \hline
      \end{tabular}
\end{table}
Among the various extensions of the SM, the $B-L$ gauge model is one of the simplest extension which has been studied in previous
 works \cite{U1X0, U1X1, U1X2, U1X3, U1X4, Khalil08, Khalil09, VienD4BL20, VienQ6BL20, S3BLGM2019} whereby the anomalies are
  canceled in different ways\footnote{The anomalies mentioned here are for continuous symmetries.
  Anomalies of non-Abelian discrete symmetries are  presented in Refs. \cite{Araki2008, Ishimori12b, Kobayashi2020}.
  The $S_4$ group is isomorphic to $(Z_2\otimes Z_2)\rtimes S_3$, and then the $Z_2$ symmetry of $S_3$ can
  be anomalous in $S_4$. Three representations $\underline{2}$, $\underline{3}$ and $\underline{1}^'$  of $S_4$
  have $\det \rho(g)=-1$ thus the odd number of $\underline{2}$ , $\underline{3}$ and $\underline{1}^'$ can
   lead to anomalies \cite{Ishimori12b, Kobayashi2020}. However, in the model consideration, there is no irreducible
    representation $\underline{1}^'$ is used and the number of irreducible representation $\underline{2}$, as well as the
    number of irreducible representation $\underline{3}$, of $S_4$ is even, therefore  $S_4$ is an anomaly-free symmetry.}.
    In this work, we improve the model proposed in Refs.\cite{Khalil08, Khalil09} thereby
     neutrino masses and various
     other phenomena involving
     leptogenesis, dark matter, etc, are satisfied. It is emphasized that the above mentioned  model
      by itself cannot
      predict the recent observed neutrino oscillation data.

It is worth mentioning that non-Abelian discrete symmetries have revealed many outstanding issues. Consequently, many of them have been applied in
explaining the observed neutrino oscillation pattern; and one of them the
 $S_{4}$ symmetry has been widely used because it provides a viable description of the observed neutrino oscillation data \cite{Altarelli2009gn, Bazzocchi2009da, Bazzocchi2009pv, Toorop2010yh, Patel2010hr, Ishimori2010fs, Dong2010lsv,Morisi2011pm,Hagedorn2011un,Altarelli2012bn,Mohapatra2012tb,BhupalDev2012nm,Varzielas2012pa,Ding2013hpa,Ding2013eca,
Campos2014zaa,VL2014, Vien2015lk, Vien2016, deAnda2017yeb,deAnda2018oik,Medeiros2019hur,Chen2019oey,VLA2019, Petcov2020, Ishimori12b, AntonioNPB20}.
However, the above mentioned models contain non minimal scalar sectors with many Higgs doublets.
Thus, it is interesting to find an alternative extension of
better explanation for
the observed neutrino oscillation data with less scalar content than previous models. In this work, we propose an alternative and improved
version of the $B-L$ model with an additional flavor symmetry group $S_4\otimes Z_{3}$ which accommodates the current neutrino oscillation
 data given in Table \ref{tab1}. In this work, all left-handed leptons are put in $\underline{3}$ while for the right-handed leptons, the first generation
  is put in $\underline{1}$ and the two others are in $\underline{2}$. The $S_4$ group contains 24 elements dispensed into five
conjugacy classes and five irreducible representations, denoted as $\underline{1}, \underline{1}^'$,
$\underline{2}, \underline{3}$, and $\underline{3}^'$. In this paper, we work in the basis where $\underline{3}$ and $\underline{3}^'$ are
real whereas $\underline{2}$ is complex. For a detailed description of $S_4$ group, the reader is referred to  Ref. \cite{Ishimori12b}.
Despite the $S_4$ symmetry has been previously studied in various works \cite{Altarelli2009gn, Bazzocchi2009da, Bazzocchi2009pv, Toorop2010yh, Patel2010hr, Ishimori2010fs, Dong2010lsv,Morisi2011pm,Hagedorn2011un,Altarelli2012bn,Mohapatra2012tb,BhupalDev2012nm,Varzielas2012pa,Ding2013hpa,Ding2013eca,
Campos2014zaa,VL2014, Vien2015lk, Vien2016, deAnda2017yeb,deAnda2018oik,Medeiros2019hur,Chen2019oey,VLA2019, Petcov2020, Ishimori12b, AntonioNPB20},
to the best of our knowledge, this symmetry has not been considered before in the $B-L$ scenario.

This paper
is arranged as follows. The model is described in Section \ref{model}. Section \ref{lepton}  is devoted to
 neutrino mass and mixing. The results of the numerical analysis are presented in Section \ref{NR} and finally, some conclusions are given in Section \ref{conclusion}.

\section{The model \label{model}}
The full symmetry of the model
is $G=G_{SM}\otimes U(1)_{B-L}\otimes S_4\otimes Z_3$ where $G_{SM}=SU(3)_C
 \otimes SU(2)_L \otimes U(1)_Y$ is the gauge group of the SM.
In this model, the first generation of right-handed lepton is put in $\underline{1}$ while the two others are put in $\underline{2}$
under $S_4$ and three generations of left-handed lepton as well as three right-handed neutrino are put in $\underline{3}$.
The model particle content  is given in Table \ref{lepcont} where $\phi, \phi^'$ and $\eta$ are $S_4$ triplets whose
 components are $SU(2)_L$ singlets and $\chi$ is one $S_4$ doublets whose components are $SU(2)_L$ singlets. Under $SU(3)_L$ symmetry, leptons and scalar fields are all put in singlet $\mathbf{1}$ whereas under $SU(2)_L$ symmetry,  $\psi_{L}$ and $H$ are put in $\mathbf{2}$ while all the others are put in $\mathbf{1}$. Therefore, the assignments under $SU(3)_C$ and $SU(2)_L$ are not presented in Table \ref{lepcont}.
\begin{table}[h]
\caption{\label{lepcont} The model  particle and scalar contents ($\al =2,3$)}
\vspace*{-0.2cm}
\begin{center}
\begin{tabular}{|c|c|c|c|c|c|c|c|c|c|} \hline
  Fields & $\psi_{L}$ &$l_{1R} (l_{\al R})$&\,\,$\nu_{R}$\,\,&\,\,$H$\,\,&\,\,$\phi$\,\,&\,\,$\phi^'$\,\,&\,\,$\chi$\,\,&\,\,$\eta$ \\ \hline
$U(1)_Y$  &  $-\fr{1}{2}$ &$-1$&$0$& $\fr{1}{2}$& $0$& $0$  & $0$ & $0$  \\
$U(1)_{B-L}$ & $-1$ &$-1$&$-1$&   $0$         & $0$& $0$  & $2$ & $2$ \\
$S_4$&  $\underline{3}$  &$\underline{1} (\underline{2})$&$\underline{3}$&$\underline{1}$& $\underline{3}$  & $\underline{3}^'$
&$\underline{2}$&$\underline{3}$  \\
$Z_3$&  $\om^2$  &$\om$&$\om^2$&$1$&$\om$& $\om$  & $\om^2$ & $\om^2$\\
\hline
\end{tabular}
\vspace*{-0.3cm}
\end{center}
\end{table}

Taking into account  that    $\bar{\psi}_{L} l_{1R}$ and $\bar{\psi}_{L} l_{\al R}$ transform,  under $G$ symmetry, as
	$(1, 2, -1/2, 0, \underline{3}, \om^2)$ and $(1, 2, -1/2, 0, \underline{3}\oplus
 \underline{3}^' , \om^2)$, respectively. Thus, we need one $SU(2)_L$ doublet $H$ and two $SU(2)_L$ singlets $\phi, \phi^'$, as presented  in Tab. \ref{lepcont},  to generate masses for the charged-leptons.

Furthermore, the neutrino masses arise from
$\bar{\psi}_{L} \nu_{R}$
and $\bar{\nu}^c_{R} \nu_{R}$ to scalars, where under $G$ symmetry, $\bar{\psi}_{L} \nu_{R}\sim (1,2,1/2, 0, \underline{1}
\oplus \underline{2}\oplus \underline{3}\oplus \underline{3}^', 1)$ and
$\bar{\nu}^c_{R} \nu_{R}\sim (1, 1,0, -2, \underline{1} \oplus \underline{2}\oplus \underline{3}\oplus \underline{3}^' , \om)$.
For the known scalars ($H, \phi, \phi^'$), under $G$ symmetry, there is only one invariant term  $(\bar{\psi}_L \nu_{R})_{\underline{1}}\widetilde{H}$
 which is responsible for generating Dirac masses for the neutrinos.  In order to
generate the realistic neutrino masses and mixings, we add
two singlets $\chi, \eta$ respectively put in $\underline{2}$
 and $\underline{3}$ under $S_4$  coupling to $\bar{\nu}^c_{L}\nu_{R}$ which are responsible for generating Majorana masses for the neutrinos.
\newline
The following Yukawa couplings are invariant under all symmetries of the model:
\bea -\mathcal{L}_{l}&=&\fr{h_1}{\La } (\bar{\psi}_{L} l_{1R})_{\underline{3}} (H\phi)_{\underline{3}}+
 \fr{h_2}{\La } (\bar{\psi}_{L} l_{\al R})_{\underline{3}} (H\phi)_{\underline{3}}
+\fr{h_3}{\La } (\bar{\psi}_{L} l_{\al R})_{\underline{3}^'} (H\phi^')_{\underline{3}^'}\crn
&+& x (\bar{\psi}_L \nu_{R})_{\underline{1}}\widetilde{H}+\fr{y}{2} (\bar{\nu}^c_R \nu_{R})_{\underline{2}}\chi
+\fr{z}{2} (\bar{\nu}^c_R \nu_{R})_{\underline{3}}\eta+H.c, \label{Lyukawa}\eea
where $\La $ is the cut-off scale of the theory, and $h_{1,2,3}$ as well as $x, y, z$ are the dimensionless
Yukawa coupling constants.

To generate the suitable
neutrino oscillation pattern, the following structure of  VEVs is chosen:
\bea
&&\langle H \rangle^T = ( 0 \hspace{0.25 cm}   v_H) , \hs \langle \phi \rangle = (\langle \phi_1 \rangle,\hspace{0.1 cm} \langle \phi_1
\rangle, \hspace{0.1 cm}\langle \phi_1 \rangle),\hspace{0.1 cm}\langle \phi_1 \rangle = v_\phi,\crn
&& \langle \phi^' \rangle = (\langle \phi^'_1 \rangle,\hspace{0.1 cm} \langle \phi^'_1 \rangle,\hspace{0.1 cm} \langle \phi^'_1 \rangle),
\hspace{0.1 cm} \langle \phi^'_1 \rangle=v_{\phi^'}, \hspace{0.1 cm}\langle \eta \rangle = (0,\hspace{0.1 cm} 0, \hspace{0.1 cm}
 \langle \eta_3 \rangle) ,\hspace{0.1 cm} \langle \eta_3 \rangle =v_\eta, \crn
&&\langle \chi \rangle =(\langle \chi_1 \rangle,\hspace{0.1 cm} \langle \chi_2 \rangle),\hspace{0.1 cm} \langle \chi_{1,2}
 \rangle =v_{\chi_{1,2}}. \label{scalarvev}
\eea
 It is to be noted that the VEVs of $\phi$ and $\phi^'$, respectively, break $S_4$ down to $S_3$ and $Z_3$ while the
VEVs of $\chi$ and $\eta$ break $S_4$ down to the Klein four group $K_4$.

From Eq. (\ref{Lyukawa}), with expansion $\phi_i=\langle \phi_i\rangle +\phi^'_i$ and $H=(H^+\hs H^0)^T$, we get the lepton flavor changing interactions:
 \bea
-\mathcal{L}_{clep}&\supset& \fr{h_1 v_{\phi}}{\La }(\bar{l}_{2L}H^0 + \bar{\nu}_{2L}H^+) l_{1R}+\fr{h_1 v_{\phi}}{\La }(\bar{l}_{3L}H^0+\bar{\nu}_{3L}H^+) l_{1R}\crn
&+&  \fr{h_2 v_{\phi}}{\La } (\bar{l}_{1L}H^0+ \bar{\nu}_{1L}H^+)l_{2R}+ \fr{h_2 v_{\phi}}{\La } (\bar{l}_{1L}H^0 + \bar{\nu}_{1L}H^+)l_{3R} \crn
&-& \fr{h_3 v_{\phi^'}}{\La }(\bar{l}_{1L}H^0 + \bar{\nu}_{1L}H^+)l_{2R}+ \fr{h_3 v_{\phi^'}}{\La }(\bar{l}_{1L}H^0 + \bar{\nu}_{1L}H^+)l_{3R} + H.c \, .
 \label{ctl3}
 \eea
Eq.\eqref{ctl3} shows that, in the case $v_{\phi^'}\simeq v_{\phi}$, the usual Yukawa couplings are proportional to 
$\frac{v_{\phi}}{ \La}$ and the lepton flavor changing processes in this model are suppressed by the factor\footnote{Here, $G_F=g^2/(4\sqrt{2}m^2_W)$ and $M_H$ is the mass scale of the heavy scalars providing the dominant contributions to the lepton flavor violation decays.} $\fr{v_{\phi}}{\La G_F^2M_H^2}$ associated with the above small Yukawa couplings and the large mass scale of the heavy scalars. For further details, the reader is referred to Refs. \cite{br1,br2,br3,br4}.
Furthermore, this model contains
only one $SU(2)_L$ Higgs doublet, therefore,
 the flavor changing neutral current processes are absent at tree level.

\section{Neutrino mass and mixing \label{lepton}}

From the Yukawa interactions in Eq. (\ref{Lyukawa}), using the tensor product of $S_4$ \cite{Ishimori12b},
together with the VEVs of $H$, $\phi$ and $\phi^'$ in Eq. (\ref{scalarvev}), the charged lepton mass terms is written as follows
 \bea
\mathcal{L}^{\mathrm{mass}}_{cl}
&=& - (\bar{l}_{1L} \hs \bar{l}_{2L}\hs \bar{l}_{3L})
M_l (l_{1R}\hs l_{2R}\hs l_{3R})^T+H.c, \eea
where
\bea M_l=\fr{v_H}{\La }\left(%
\begin{array}{ccc}
  h_1 v_\phi& h_2 v_\phi-h_3 v_{\phi^'} & h_2 v_\phi+ h_3 v_{\phi^'}  \\
  h_1 v_\phi&  \om^2 (h_2 v_\phi-h_3 v_{\phi^'})&  \om (h_2 v_\phi+ h_3 v_{\phi^'})\\
 h_1 v_\phi&  \om (h_2 v_\phi-h_3 v_{\phi^'})& \om^2 (h_2 v_\phi+ h_3 v_{\phi^'})\\
\end{array}%
\right).\label{Mltq}\eea
The matrix $M_l$  is diagonalized as $ U^\dagger_l M_l U_r= \mathrm{diag}(m_e, \, m_\mu , \,m_\tau)$, where
\bea U^\dagger_L&=&\fr{1}{\sqrt{3}}\left(%
\begin{array}{ccc}
  1 &\,\,\, 1 &\,\,\, 1 \\
  1 &\,\,\, \om &\,\,\, \om^2 \\
  1 &\,\,\, \om^2 &\,\,\, \om \\
\end{array}%
\right),\hs U_R=I_{3\times 3}, \hs \om=e^{i2\pi/3},  \label{Uclep}\\
m_e &=&\fr{\sqrt{3} h_1 v_H v_{\phi}}{\La },\hs  m_{\mu,\tau}=\fr{\sqrt{3}v_H}{\La }
\left(h_2v_{\phi} \mp h_3v_{\phi^'}\right). \label{memt}\eea
The left-handed mixing matrix $U_L$ is non trivial in
our model and hence 
will contribute to the leptonic mixing matrix. The
Eq. (\ref{memt}) shows that $m_{\mu}$ and $m_{\tau}$ are differentiated
by $\phi^'$. This is the motive
why $\phi^'$ is additionally introduced to $\phi$ in the charged-lepton sector.
Now, comparing the result in Eq (\ref{memt}) with the best fit values for the masses of charged-leptons taken
from Ref. \cite{PDG2020}, $m_e \simeq 0.511 \, \textrm{MeV}$, $\, m_\mu \simeq 105.66 \, \textrm{MeV}$,
$\, m_\tau \simeq 1776.86 \,\textrm{MeV}$, we find the relations $\fr{h_1 v_{H} v_\phi}{\La }= 0.295\,
 \mathrm{MeV},\hs \fr{h_2 v_{H} v_\phi}{\La }= 543\, \mathrm{MeV}, \hs
 \fr{h_3 v_{H} v_{\phi^'}}{\La }= 482\, \mathrm{MeV}$, i.e., $h_1:h_2:h_3\sim 1.00: 1.840\times 10^3: 1.634\times 10^3$.

Regarding the neutrino sector, from the Yukawa terms in Eq. (\ref{Lyukawa}) and using the tensor product of $S_4$ \cite{Ishimori12b},
 the Yukawa Lagrangian invariant under $\mathrm{G}$ symmetry in neutrino sector reads
\bea
 -\mathcal{L}_{\nu}&=& x \left(\bar{\psi}_{1L} \nu_{1R} +\bar{\psi}_{2L} \nu_{2R} +\bar{\psi}_{2L} \nu_{3R} \right)\widetilde{H} \crn
&+&\fr{y}{2}  \left[(\chi_1+\chi_2)\bar{\nu}^c_{1R}\nu_{1R} +\om(\om \chi_1+ \chi_2)\bar{\nu}^c_{2R}\nu_{2R}
+\om(\chi_1+\om \chi_2)\bar{\nu}^c_{3R}\nu_{3R}\right]\crn
&+&\fr{z}{2} \left[(\bar{\nu}^c_{2R}\nu_{3R}+\bar{\nu}^c_{3R}\nu_{2R})\eta_1+(\bar{\nu}^c_{3R}\nu_{1R}+\bar{\nu}^c_{1R}\nu_{3R})\eta_2
+(\bar{\nu}^c_{1R}\nu_{2R}+\bar{\nu}^c_{2R}\nu_{1R})\eta_3 \right]\crn
&+& H.c.\label{Lny}\eea
After symmetry breaking, the mass Lagrangian for the
neutrinos gets the following form:
\bea -\mathcal{L}^{mass}_{\nu}&=&\fr 1 2
\bar{\chi}^c_L M_\nu \chi_L+ H.c,\label{nm}\eea where \bea
\chi_L&=& \left(\nu_L \hs
  \nu^c_R \right)^T,\hs M_\nu = \left(%
\begin{array}{cc}
  0 & M^T_D \\
  M_D & M_R \\
\end{array}%
\right), \label{MnuLDR}\\
 \nu_L&=&(\nu_{1L}\hs\nu_{2L}\hs\nu_{3L})^T,\hs
\nu^c_R=(\nu^c_{1R}\hs \nu^c_{2R}\hs\nu^c_{3R})^T, \nn  \eea and the mass matrices $M_D, M_R$ take the following forms
\bea M_D= \mathrm{diag}(a_D, \hspace{0.15cm} a_D,  \hspace{0.15cm} a_D),
\, M_R=
\left(%
\begin{array}{ccc}
a_{1R}+ a_{2R} & a_{R}                  & 0 \\
 a_{R}         &\om (\om a_{1R}+a_{2R}) & 0 \\
 0             &  0                     & \om (a_{1R}+\om a_{2R}) \\
\end{array}%
\right), \label{MDR}\eea
where \bea
 a_D=x v^*_{\phi},  \,\, a_{1,2 R}= y v_{\chi_{1,2}},\,\, a_{R}=z v_\eta.  \label{aa123DR}\eea
In seesaw mechanism, the effective neutrino mass matrix is given by
 \bea
{M}_{\mathrm{eff}}&=&-M_D^T{M}_R^{-1}M_D=\left(%
\begin{array}{ccc}
  A_1+i A_2 & A_7+i A_8 & 0 \\
  A_7+i A_8 & A_3+i A_4 & 0 \\
  0         & 0 & A_5+i A_6 \\
\end{array}%
\right), \label{Meff}\eea
where
\bea
&& A_1 A_0 = -\left[2 (a_{1R}^3 + a_{2R}^3) + (a_{1R} + a_{2R}) a_R^2\right]a_D^2, \crn
&& A_2 A_0=\sqrt{3} (a_{2R}-a_{1R}) a_D^2 a_R^2, \crn
&& A_3 A_0=(a_{1R} + a_{2R}) a_D^2 [(a_{1R} + a_{2R})^2 + 2 a_R^2], \crn
&& A_4 A_0=\sqrt{3} (a_{2R} - a_{1R}) (a_{1R} + a_{2R})^2 a_D^2,\crn
&& A_5 =\fr{(a_{1R} + a_{2R}) a_D^2}{2\left(a_{1R}^2 - a_{1R} a_{2R} + a_{2R}^2\right)}, \hs
A_6=\fr{\sqrt{3}}{2}\fr{(a_{2R}-a_{1R}) a_D^2}{a_{1R}^2 - a_{1R} a_{2R} + a_{2R}^2}, \crn
&& A_7 A_0 = -a_D^2 a_R \left[(a_{1R} + a_{2R})^2 + 2 a_R^2\right], \crn
&& A_8 A_0=\sqrt{3} (a_{1R}^2-a_{2R}^2) a_D^2 a_R,\crn
&& A_0= 2 \left[a_{1R}^4 + a_{1R}^3 a_{2R} +
   a_{1R} a_{2R}^3 + a_{2R}^4 + (a_{1R} + a_{2R})^2 a_R^2 + a_R^4\right].\label{ABCM}
\eea
Let us firstly define a Hermitian matrix $\mathcal{M}^2$, given by
\bea
\mathcal{M}^2&=& {M}_{\mathrm{eff}} {M}_{\mathrm{eff}}^+ = \left(%
\begin{array}{ccc}
  a_0 & d_0+ig_0 &\hs 0 \\
  d_0-ig_0 &  b_0& 0\\
  0\hs & 0 & c_0\\
\end{array}%
\right), \label{Msq}
\eea
where
\bea
&&a_0= a_1^2 + a_2^2 + a_7^2 + a_8^2 + 2 a_1 a_2 \sin(\al _1-\al _2) + 2 a_7 a_8 \sin(\al _7-\al _8), \crn
&&b_0 = a_3^2 + a_4^2 + a_7^2 + a_8^2 + 2 a_3 a_4 \sin(\al _3-\al _4)+ 2 a_7 a_8 \sin(\al _7-\al _8), \crn
&&c_0 = a_5^2 + a_6^2 + 2 a_5 a_6 \sin(\al _5-\al _6),\crn
&&d_0=a_7 [a_1 \cos(\al _1-\al _7)- a_2 \sin(\al _2-\al _7)+ a_3 \cos(\al _3-\al _7)+a_4 \sin(\al _4-\al _7)] \crn
&&\hspace{0.45cm}+\, a_8 [a_1 \sin(\al _1-\al _8)+a_2 \cos(\al _2-\al _8) + a_3 \sin(\al _3-\al _8)+
a_4 \cos(\al _4-\al _8)], \crn
&& g_0= a_7[a_1 \sin(\al _1-\al _7)+a_2\cos(\al _2-\al _7) -a_3 \cos(\al _3-\al _7)-a_4 \sin(\al _4-\al _7)]\crn
&& \hspace{0.5cm} -\, a_8[a_1 \cos(\al _1-\al _8)-a_2 \sin(\al _2-\al _8)-a_3 \cos(\al _3-\al _8)+
a_4\sin(\al _4-\al _8)],\label{abcdg0}
\eea
with $a_i =|A_i|$ and $\al _i\, (i=1\div 8)$ are the arguments of $A_i$.
\newline
The squared-mass matrix $\mathcal{M}^2$  in Eq. (\ref{Msq}) owns
three exact eigenvalues
\bea
m_1&=&\ka _1-\ka _2,\hs m_2=c_0,\hs m_{3}=\ka _1+\ka _2, \label{m1m2m3}
\eea
with
\bea
2 \ka _1 = a_0 + b_0, \hs 2\ka _2 =\sqrt{(a_0 - b_0)^2 + 4 (d_0^2 + g_0^2)},\label{kappa12}
\eea
and the corresponding mixing matrix is
\bea
U=\left(%
\begin{array}{ccc}
 \mathrm{cos} \theta & 0 &\hs -\mathrm{sin} \theta \, e^{i \al } \\
  0 & 1 & 0 \\
 \mathrm{sin} \theta  e^{-i \al }\hs & 0 & \mathrm{cos} \theta\\
\end{array}%
\right),
\eea
with
\bea
 \al &=&-i\ln\left(-\fr{d_0 +i g_0}{\sqrt{d^2_0 + g^2_0}}\right),\hs \theta=\arcsin\left(\fr{1}{K^2+1}\right),\label{alphatheta}\\
K&=&\fr{b_0-m_1}{\sqrt{d_0^2 + g_0^2}}=\fr{m_3-a_0}{\sqrt{d_0^2 + g_0^2}}.
\eea
The sign of $\De  m^2_{31}$ plays a pivotal role in form of the neutrino mass hierarchy. In the NH, $m_1\ll m_{2}\sim m_{3}$ thus the lightest neutrino mass is $m_1$ while in the IH, $m_3\ll m_1\sim m_2$
thus the lightest neutrino mass is $m_3$ \cite{PDG2020}. The neutrino mass matrix ${M}_{\mathrm{eff}}$ in Eq. (\ref{Meff}) is diagonalized as
\be
U_{\nu }^+ \mathcal{M}^2 U_{\nu }=\left\{
\begin{array}{l}
\left(
\begin{array}{ccc}
m_1 & 0 & 0 \\
0 & m_{2} & 0 \\
0 & 0 & m_{3}%
\end{array}%
\right) ,\hspace{0.1cm} U_{\nu }=\left(%
\begin{array}{ccc}
 \mathrm{cos} \theta & 0 &\hs -\mathrm{sin} \theta \, e^{i \al } \\
 \mathrm{sin} \theta  e^{-i \al }\hs & 0 & \mathrm{cos} \theta\\
 0 & 1 & 0 \\
\end{array}%
\right) \hspace{0.2cm}\mbox{for NH,}\ \  \\
\left(
\begin{array}{ccc}
m_{3} & 0 & 0 \\
0 & m_{2} & 0 \\
0 & 0 & m_1
\end{array}%
\right) ,\hspace{0.1cm} U_{\nu }=\left(%
\begin{array}{ccc}
 \mathrm{sin} \theta \, e^{i \al } \hs& 0 &\hs \mathrm{cos} \theta  \\
-\mathrm{cos} \theta \hs & 0 & \hs\mathrm{sin} \theta  e^{-i \al }\\
 0 & 1 & 0 \\
\end{array}%
\right) \hspace{0.2cm}\mbox{for IH,}%
\end{array}%
\right.  \label{Unu}
\ee
where $m_{2,3}$ and $\al , \theta$ are given in Eqs. (\ref{m1m2m3}) and (\ref{alphatheta}), respectively.
The corresponding leptonic mixing matrix is
\be
U_{lep}=U_{L}^{\dag} U_{\nu }=\left\{
\begin{array}{l}
\fr{1}{\sqrt{3}}\left( \begin{array}{ccc}
\cos\theta + \sin\theta. e^{-i \al }      &1 & \cos\theta - \sin\theta. e^{i \al }  \\
\cos\theta + \om \sin\theta. e^{-i \al }\hs &\om^2 &\hs \om\cos\theta - \sin\theta. e^{i \al }  \\
\cos\theta + \om^2 \sin\theta. e^{-i \al }&\om  & \om^2\cos\theta - \sin\theta. e^{i \al } \\
\end{array}\right) \hspace{0.2cm}\mbox{for \, NH},  \label{Ulep}  \\
\fr{1}{\sqrt{3}}\left( \begin{array}{ccc}
-\cos\theta + \sin\theta. e^{i \al }     &1 & \cos\theta + \sin\theta. e^{-i \al } \\
-\om\cos\theta + \sin\theta. e^{i \al } &\om^2 &\hs \cos\theta + \om \sin\theta. e^{-i \al }  \\
-\om^2\cos\theta + \sin\theta. e^{i \al } &\hs\om & \cos\theta + \om^2 \sin\theta. e^{-i \al } \\
\end{array}\right) \hspace{0.2cm}\mbox{for \, IH}.
\end{array}%
\right.
\ee
 In the three-neutrino scheme, lepton mixing angles can be defined as:
\bea s_{13}^2=\left| \mathrm{U}_{e 3}\right|^2, \,\, t_{12}^2 = \left|\fr{\mathrm{U}_{e 2}}{\mathrm{U}_{e 1}}\right|^2,\,\,
t_{23}^2=\left|\fr{\mathrm{U}_{\mu 3}}{\mathrm{U}_{\tau 3}}\right|^2,\label{t12t23s13}\eea
where $t_{12}=s_{12}/c_{12}$, $t_{23}=s_{23}/c_{23}$, $c_{ij}=\cos \theta_{ij}$, $s_{ij}=\sin \theta_{ij}$ with
$\theta_{ij}$ are neutrino mixing angles.

Combining the standard parametrization of the lepton mixing
 matrix \cite{Pontecorvo1, Pontecorvo2, Jcp, Maki, Rodejohann, Jarlskog1, Jarlskog2, Jarlskog3} and the
  expression (\ref{Ulep}), the Jarlskog invariant 
  constraining  the size of CP violation in lepton sector,
  is determined as \cite{Jarlskog1, Jarlskog2, Jarlskog3}:
\bea
J_{CP}&=&\mathrm{Im} (U_{12} U_{23} U^*_{13} U^*_{22})=\left\{
\begin{array}{l}
-\fr{\cos(2\theta)}{6\sqrt{3}} \hspace{0.3cm}\mbox{for \, NH},    \\
\hspace{0.3 cm} \fr{\cos(2\theta)}{6\sqrt{3}}  \hspace{0.25cm}\,\mbox{for \, IH}.
\end{array}%
\right. \label{Jm}
\eea
Expressions in (\ref{Ulep}), (\ref{t12t23s13}) and (\ref{Jm}) yield the following relations:
\bea
&& \cos\theta =\left\{
\begin{array}{l}
\sqrt{\fr{1}{2} - 3 \sqrt{3} J_{CP}} \hspace{0.3cm}\mbox{for \, NH},    \\
\sqrt{\fr{1}{2} + 3 \sqrt{3} J_{CP}} \hspace{0.25cm}\,\mbox{for \, IH}.
\end{array}%
\right. , \label{costheta}\\
&&\cos\al =\left\{
\begin{array}{l}
\fr{1-3 s_{13}^2}{\sqrt{1 - 108 J^2_{CP}}} \hspace{0.3cm}\mbox{for \, NH},    \\
\fr{-1+3 s_{13}^2}{\sqrt{1 - 108 J^2_{CP}}}  \hspace{0.25cm}\,\mbox{for \, IH}.
\end{array}%
\right., \label{cosalpha}\\
&&\sin\de  =-\fr{1}{2}\sqrt{4- \mathcal{O}(s^2_{13}, s^2_{23})} \hspace{0.3cm}\mbox{for \,
both NH and IH}, \label{sds13s23}\\
&&\mathcal{O}(s^2_{13}, s^2_{23}) =\fr{(1 - 2 s_{13}^2)^2 (1 - 2 s_{23}^2)^2}{s_{13}^2 s_{23}^2(2 -
3 s_{13}^2) (1 - s_{23}^2)}, \label{small}\\
&&t^2_{12} =\fr{1}{2 - 3 s_{13}^2} \hspace{0.3cm}\mbox{for \, both NH and IH}. \label{s13t12}\eea
Since $s^2_{13}$ is a very small positive number and $s^2_{23}$ is very close to $\fr{1}{2}$, we can
 approximate that $\mathcal{O}(s^2_{13}, s^2_{23}) \ll 1$.  Thus, from Eqs. (\ref{s13t12})--(\ref{sds13s23}) we can approximate
\bea
&& t^2_{12}> \fr{1}{2}, \hs \sin\de  \simeq  -1 + \fr{\mathcal{O}(s^2_{13}, s^2_{23})}{8} <0. \label{approxi}
\eea
From Eqs. (\ref{m1m2m3}), (\ref{Unu}) and (\ref{Ulep}), one can determine the effective neutrino masses
 governing the beta decay ($m_{\beta }$) and neutrinoless double beta decay ($\langle m_{ee}\rangle$),
 which can in principle determine the absolute neutrino mass scale \cite{betdecay2,betdecay4,betdecay5}
\bea
m_{\beta }= \sqrt{\sum^3_{i=1} \left|U_{ei}\right|^2 m_i^2}
,\hs \langle m_{ee}\rangle = \left| \sum^3_{i=1} U_{ei}^2 m_i \right|,
\eea
   where $m_{i}\, (i=1,2,3)$ are the masses of three active neutrinos defined from Eqs. (\ref{m1m2m3}) and (\ref{Unu})
   while $U_{ei}$ are the elements of $U_{\mathrm{PMNS}}$ determined from Eq.(\ref{Ulep}).
\section{\label{NR} Numerical analysis}
\begin{figure}[h]
\begin{center}
\vspace{-0.75 cm}
\hspace{-5.0 cm}
\includegraphics[width=0.635\textwidth]{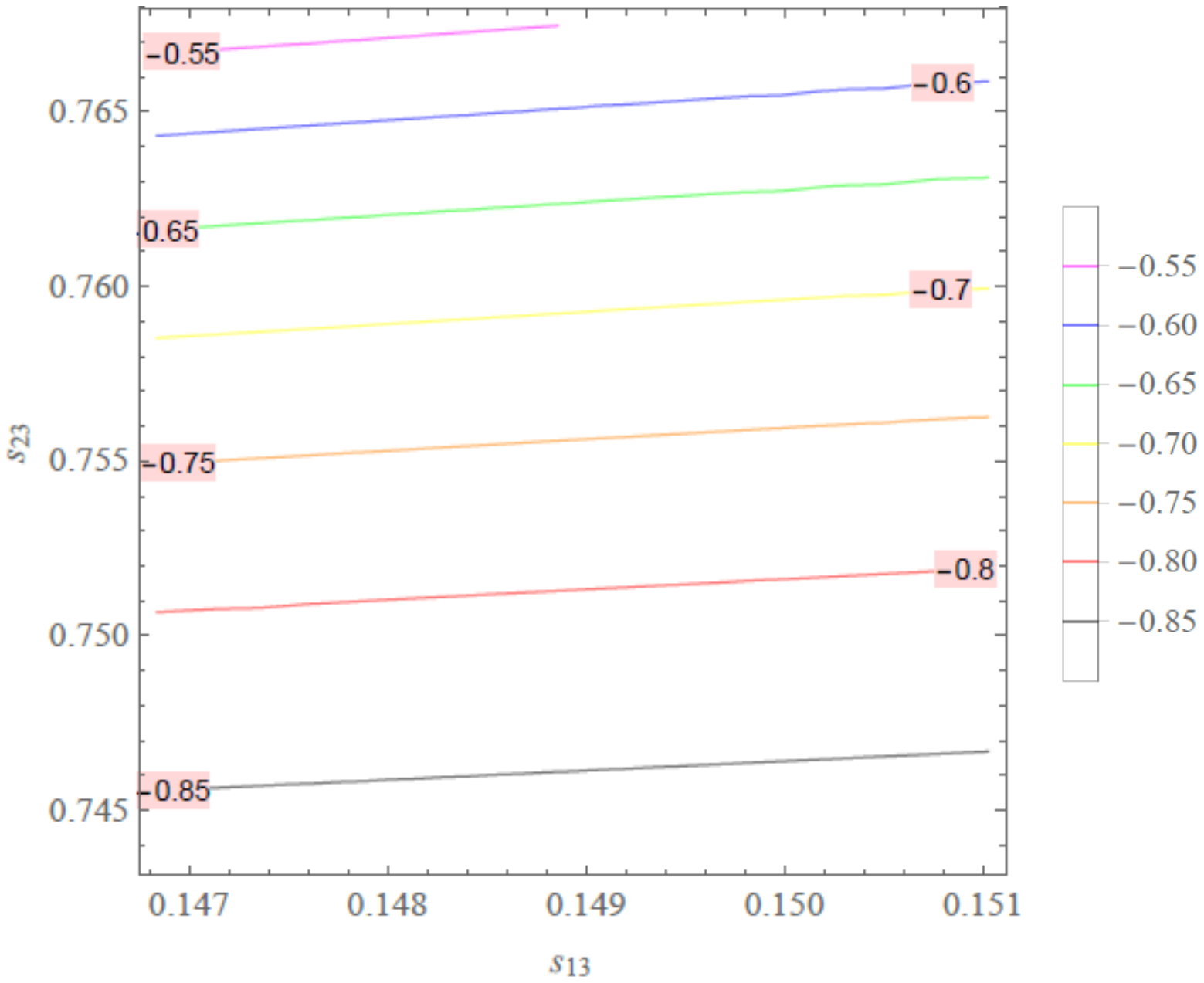}\hspace{-1.75 cm}
\includegraphics[width=0.635\textwidth]{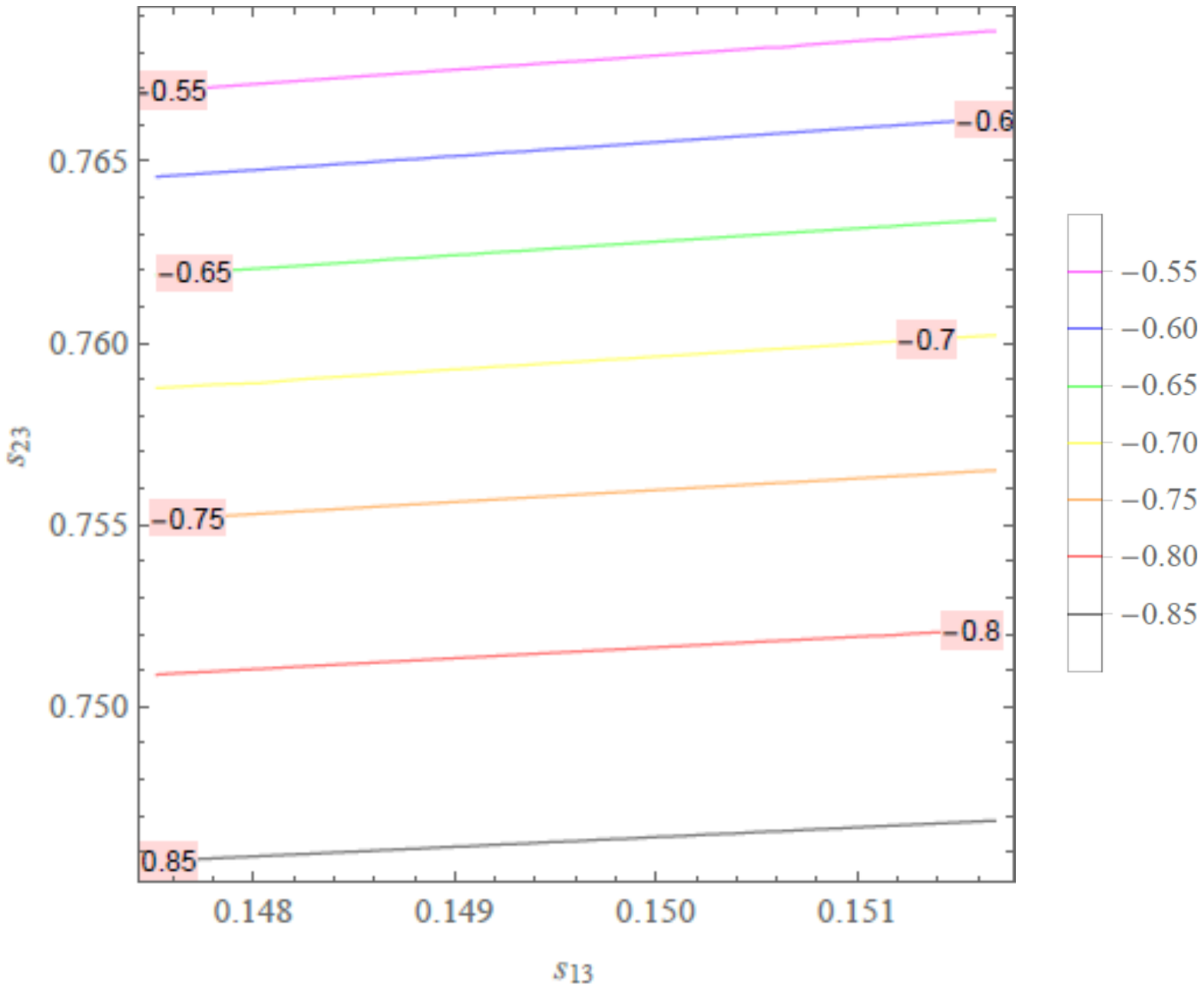}\hspace*{-5.5 cm}
\end{center}
\vspace{-7.0 cm}
\caption[The contour plot $\sin\de $ as a function of the $s_{13}$ and $s_{23}$ in the $1\si $ range of the best-fit
value taken from Ref. \cite{Esteban2020}, i.e., $s_{13}\in (0.1468,\, 0.1510)$ and $s_{23}\in (0.7436,\, 0.7675)$ for NH (left panel),
 and $s_{13}\in (0.1475,\, 0.1517)$ and $s_{23}\in (0.7457,\, 0.7688)$ for IH (left panel).]{The contour plot $\sin\de $ as
 a function of the $s_{13}$ and $s_{23}$ in the $1\si $ range of the best-fit value taken from
 Ref. \cite{Esteban2020}, i.e., $s_{13}\in (0.1468,\, 0.1510)$ and $s_{23}\in (0.7436,\, 0.7675)$ for NH (left panel),
 and $s_{13}\in (0.1475,\, 0.1517)$ and $s_{23}\in (0.7457,\, 0.7688)$ for IH (left panel).}
\label{sindelta}
\end{figure}

\begin{figure}[ht]
\begin{center}
\vspace{0.5 cm}
\hspace{-0.5 cm}
\includegraphics[width=0.475 \textwidth]{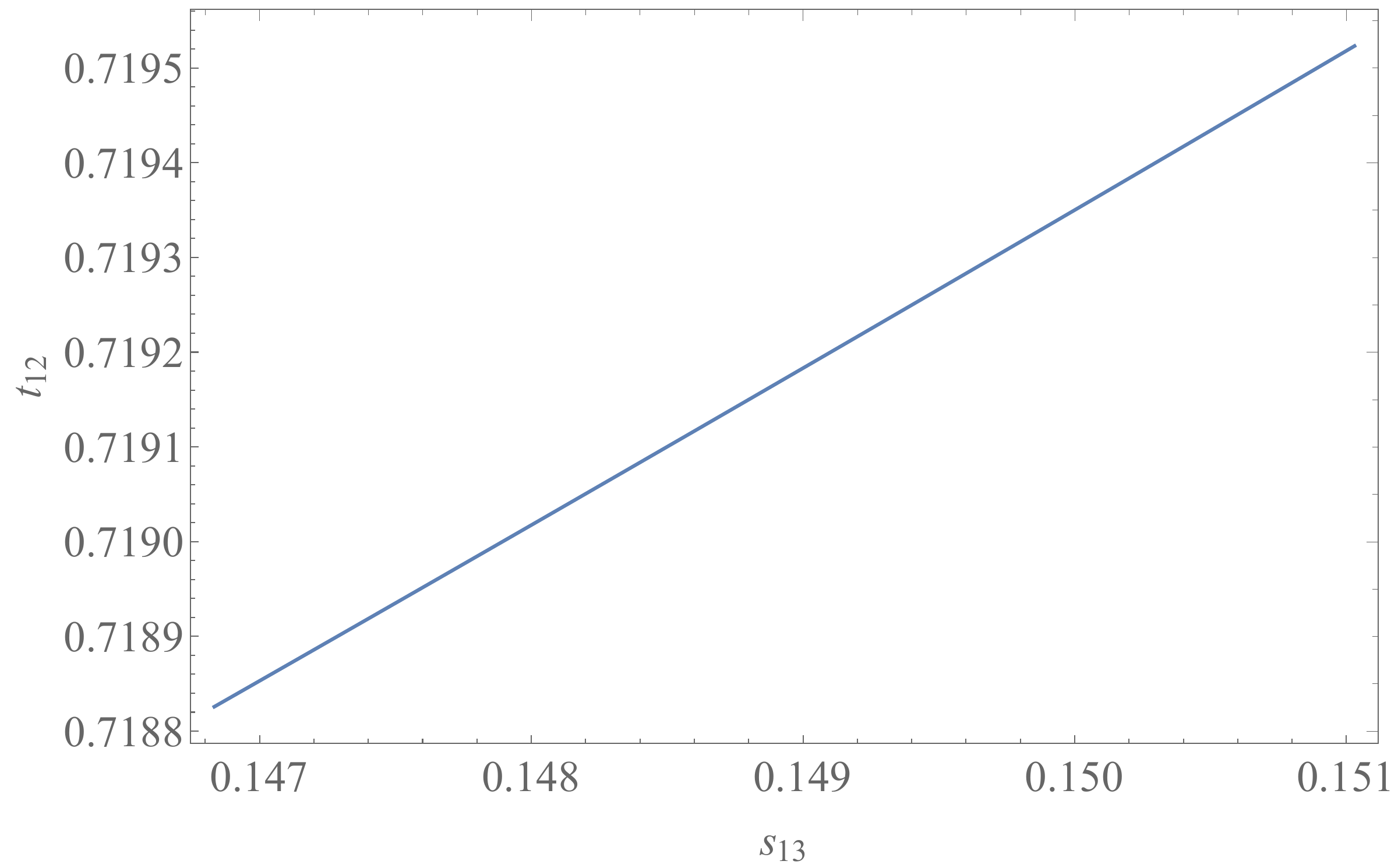}\hspace{-0.15 cm}
\includegraphics[width=0.475 \textwidth]{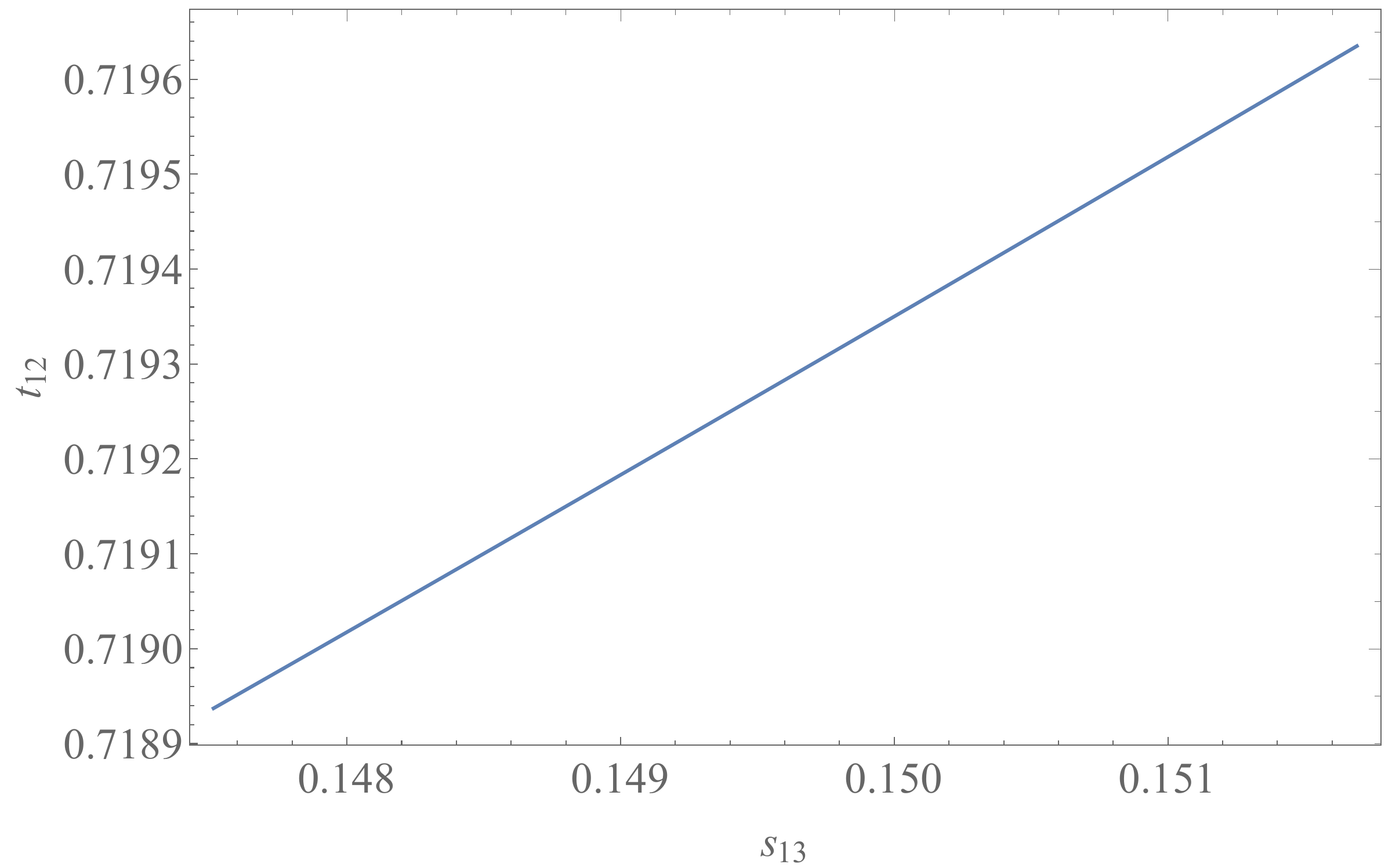}\hspace{-1.0 cm}
\end{center}
\vspace{-0.625 cm}
\caption{$t_{12}$ versus $s_{13}$ with $s_{13}\in (0.1468,\, 0.1510)$ for NH (left panel) and $s_{13}\in (0.1475,\, 0.1517)$
for IH (right panel) for the $1\si $ range of the best-fit value taken from Ref. \cite{Esteban2020}.}
\label{t12s13}
\vspace{0.35 cm}
\end{figure}

At $1\, \si $ range\footnote{Here, numbers are displayed with 4 precise digits to the right of the decimal}
 \cite{Esteban2020}, $s_{13}\in (0.1468,\, 0.1510)$ and $s_{23}\in (0.7436,\, 0.7675)$ for NH while $s_{13}\in (0.1475,\, 0.1517)$
 and $s_{23}\in (0.7457,\, 0.7688)$ for IH, thus, from  Eqs. (\ref{costheta}) and (\ref{cosalpha}) we can  find out the range of values
 of $\cos\theta$ and $\cos\al $ as plotted in Figs. \ref{costheta}, \ref{cosalpha} and \ref{sindelta}, respectively.
 On the other hand, since $s^2_{13}$ is a very small and $s^2_{23}$ is very close to $\fr{1}{2}$, we can assume
 $\mathcal{O}(s^2_{13}, s^2_{23}) \ll 1$. Thus, from Eqs. (\ref{sds13s23}) and (\ref{small}) we can approximate
\bea
\sin\de  \simeq  -1 + \fr{\mathcal{O}(s^2_{13}, s^2_{23})}{8} <0. \label{approxi}
\eea
Fig. \ref{sindelta} shows that, in $1\, \si $ range of the best-fit value taken from Ref.\cite{Esteban2020},  the  range of the Dirac CP violating phase is defined as
\bea
\sin\de  \in \left\{
\begin{array}{l}
(-0.8664, -0.5688) \hspace{0.3cm}\mbox{for \, NH},    \\
(-0.8509, -0.5462) \hspace{0.25cm}\,\mbox{for \, IH},
\end{array}%
\right. \label{sd1}
\eea
i.e.,
\bea
\de  \in \left\{
\begin{array}{l}
(299.96,\, 325.33)^\circ \hspace{0.3cm}\mbox{for \, NH},    \\
(301.70,\, 326.90)^\circ \hspace{0.25cm}\,\mbox{for \, IH}.
\end{array}%
\right. \label{sd2}
\eea
In the case $\theta_{23}$ takes its maximal values $(\theta_{23}=\fr{\pi}{4})$, $\mathcal{O}(s^2_{13}, s^2_{23})=0$
and $\sin\de =-1$,  the model predicts the cobimaximal mixing pattern \cite{Fukuura1999, Miura2000, Ma2002, He2015, Ma2015, Ma2016, Ma2017EPL, Ma2017PLB, Grimus2017, Ma2019EPJC, Ma2019NPB, Antonio2020}: $\theta_{13} \neq 0,\, \theta_{23} =\fr \pi 4$
 and $\de  =-\fr \pi 2$\ .
 Furthermore, Eq. (\ref{s13t12}) implies $t_{12}\in (0.7188,\, 0.7195)$ for NH and $t_{12}\in (0.7189,\, 0.7196)$
  for IH in $1\si $ range of $s_{13}$ of the best-fit value taken  from Ref. \cite{Esteban2020} which is plotted in Fig. \ref{t12s13}.
  These intervals of $\theta_{12}$
belong to $3\, \si $ range of the best-fit value.

\begin{figure}[h]
\begin{center}
\vspace{-0.5 cm}
\hspace{-5.0 cm}
\includegraphics[width=0.635\textwidth]{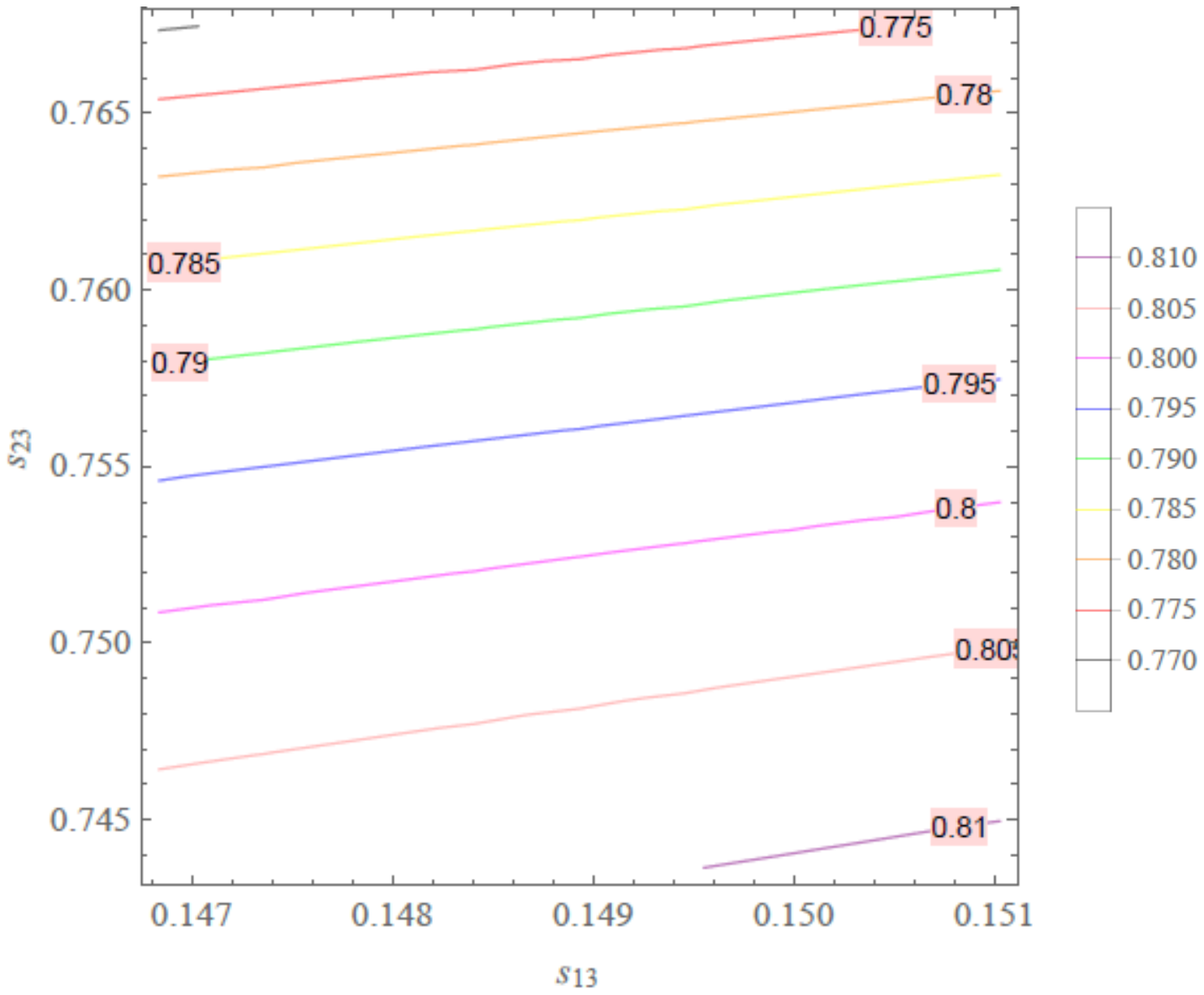}\hspace{-3.1 cm}
\vspace{-0.5 cm}
\includegraphics[width=0.635\textwidth]{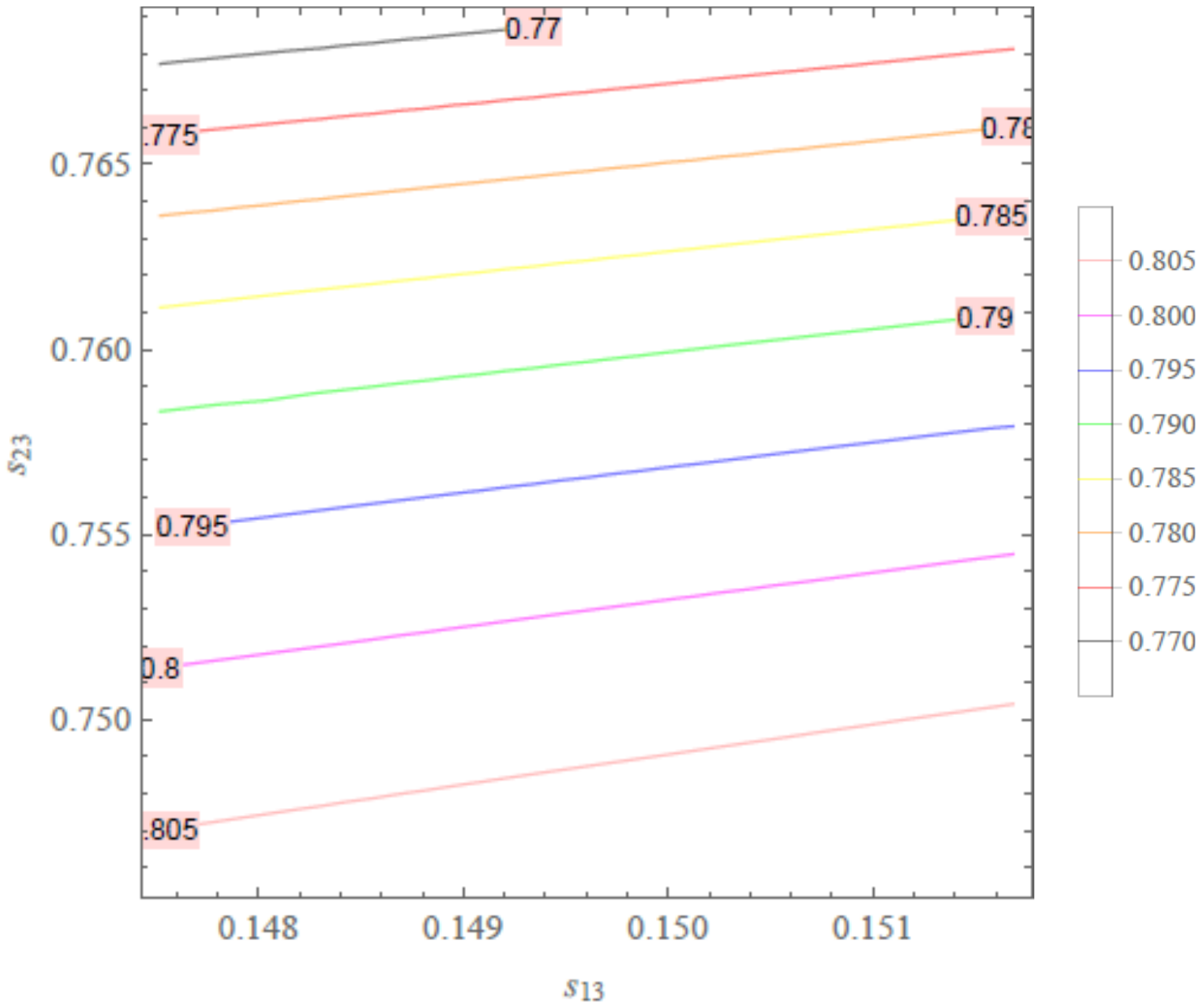}\hspace*{-5.5 cm}
\end{center}
\vspace{-6.75 cm}
\caption[The contour plot $\cos\theta$ as a function of the $s_{13}$ and $s_{23}$ in the $1\si $ range of the best-fit
 value taken from Ref. \cite{Esteban2020}, i.e., $s_{13}\in (0.1468,\, 0.1510)$ and $s_{23}\in (0.7436,\, 0.7675)$ for NH (left panel),
 and $s_{13}\in (0.1475,\, 0.1517)$ and $s_{23}\in (0.7457,\, 0.7688)$ for IH (left panel).]{The contour plot $\cos\theta$
 as a function of the $s_{13}$ and $s_{23}$ in the $1\si $ range of the best-fit value taken
 from Ref. \cite{Esteban2020}, i.e., $s_{13}\in (0.1468,\, 0.1510)$ and $s_{23}\in (0.7436,\, 0.7675)$
  for NH (left panel), and $s_{13}\in (0.1475,\, 0.1517)$ and $s_{23}\in (0.7457,\, 0.7688)$ for IH (right panel).}
\label{costheta}
\end{figure}
\begin{figure}[h]
\begin{center}
\vspace{-0.75 cm}
\hspace{-6.0 cm}
\includegraphics[width=0.675\textwidth]{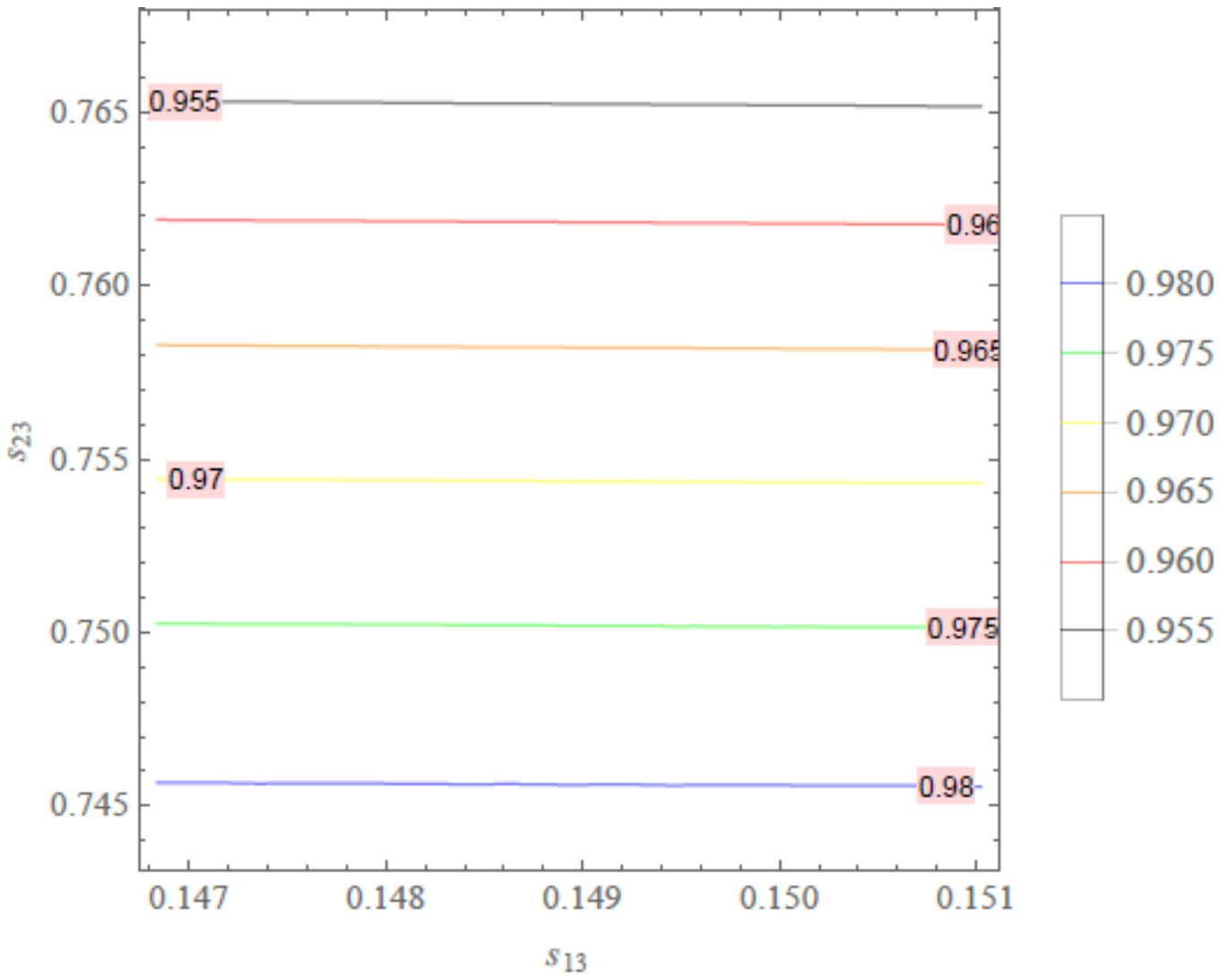}\hspace{-3.75 cm}
\includegraphics[width=0.675\textwidth]{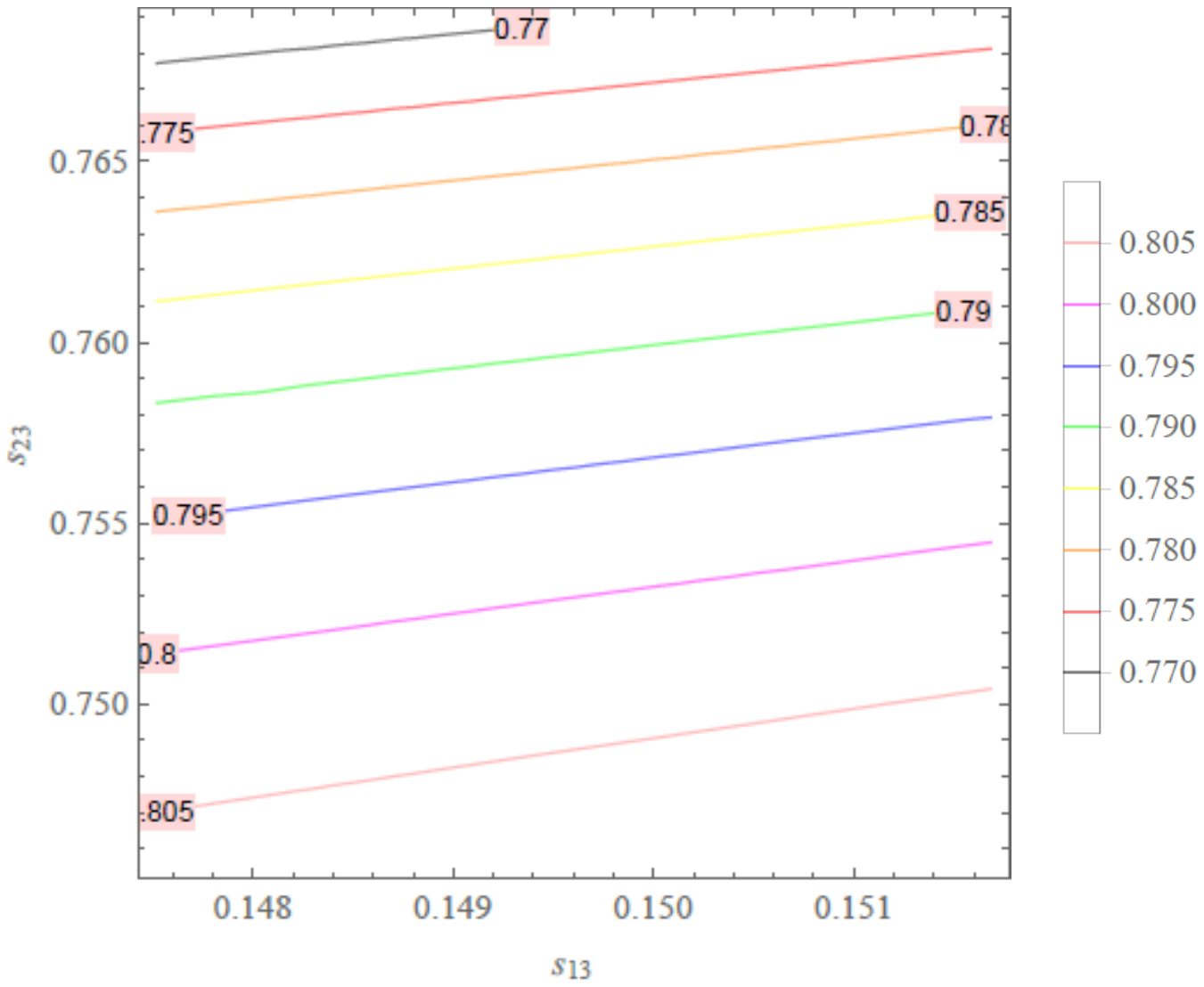}\hspace*{-7.5 cm}
\end{center}
\vspace{-8.5 cm}
\caption[The contour plot $\cos\al $ as a function of the $s_{13}$ and $s_{23}$ in the $1\si $ range of the best-fit value
taken from Ref. \cite{Esteban2020}, i.e., $s_{13}\in (0.1468,\, 0.1510)$ and $s_{23}\in (0.7436,\, 0.7675)$ for NH (left panel),
and $s_{13}\in (0.1475,\, 0.1517)$ and $s_{23}\in (0.7457,\, 0.7688)$ for IH (left panel).]{The contour plot $\cos\al $ as
a function of the $s_{13}$ and $s_{23}$ in the $1\si $ range of the best-fit value taken from
Ref. \cite{Esteban2020}, i.e., $s_{13}\in (0.1468,\, 0.1510)$ and $s_{23}\in (0.7436,\, 0.7675)$ for NH (left panel),
and $s_{13}\in (0.1475,\, 0.1517)$ and $s_{23}\in (0.7457,\, 0.7688)$ for IH (left panel).}
\label{cosalpha}
\end{figure}

Fig. \ref{sindelta} shows that, in our model, $\sin\de \in (-0.851,\, -0.546)$, i.e., $\de^\circ\in (300.0,\, 325)$ for NH and $\sin\delta\in (-0.866,\, -0.569)$, i.e., $\delta^\circ\in (301.679,\, 326.907)$ for IH. Beside that, at $1\si $ range\cite{Esteban2020}, $\delta^\circ_{CP} \in (173, 224)$ for NH and $\delta^\circ_{CP} \in (252, 308)$ for IH while at $3\si $ range \cite{Esteban2020}, $\delta^\circ_{CP} \in (120, 369)$ for NH and $\delta^\circ_{CP} \in (193, 352)$ for IH. Hence, this model predicts the Dirac CP violating phase ($\delta$) in which for NH $\delta$ belongs
to $3\, \sigma$ range and for IH $\delta$ belongs to $1\, \sigma$ range of the best-fit value taken from \cite{Esteban2020}.

In order to fix the parameters, we should deal
with the central values given in Table \ref{tab1}.
For NH, taking the central values of $\theta_{23}$ and $\theta_{13}$
 as shown in Tab. \ref{tab1}, $s_{23}=0.757,\, s_{13}=0.149$ and for IH, taking the central values of $\theta_{13}$, $s_{13}=0.1496$ and $s_{23}=0.7517$ which belongs to $1 \sigma$ range of the best-fit value taken from Ref. \cite{Esteban2020},  we get
\bea
\theta&=& \left\{
\begin{array}{l}
37.47^\circ \hspace{0.3cm}\mbox{for \, NH},    \\
53.28^\circ\hspace{0.25cm}\,\mbox{for \, IH},
\end{array}%
\right. \hspace{0.675cm} \al =\left\{
\begin{array}{l}
14.84^\circ\hspace{0.3cm}\mbox{for \, NH},    \\
166.70^\circ\hspace{0.25cm}\,\mbox{for \, IH}
\end{array}%
\right. \crn
\de &=&\left\{
\begin{array}{l}
312.91^\circ \hspace{0.3cm}\mbox{for \, NH},    \\
307.03^\circ \hspace{0.25cm}\,\mbox{for \, IH},
\end{array}%
\right. \hspace{0.5cm} \theta_{12}=\left\{
\begin{array}{l}
35.72^\circ \hspace{0.3cm}\mbox{for \, NH},    \\
35.73^\circ\hspace{0.25cm}\,\mbox{for \, IH}.
\end{array}%
\right.
\label{3}
\eea
\newline
Substitution of (\ref{3}) into (\ref{Ulep}) yields an explicit form of
the leptonic mixing matrix
\bea
U_{lep}=\left\{
\begin{array}{l}
\left(%
\begin{array}{ccc}
\,0.7977-0.08993i      & 0.5774     &0.1188-0.08993i  \\
\,0.3664 + 0.339 i &\hs -0.2887 - 0.50i  \hs &-0.5686+ 0.3069i  \\
\,0.2106-0.2490i      &\hs -0.2887 + 0.50i \hs &-0.5686-0.4868i  \\
\end{array}%
\right) \hspace{0.3cm}\mbox{for \, NH},    \\
\left(%
\begin{array}{ccc}
-0.7956+ 0.1063i & 0.5774     &-0.1053 - 0.1063i  \\
-0.2779-0.1926i &\hs -0.2887 - 0.50i  \hs &0.6624-0.3369i  \\
-0.2779+0.4052i &\hs -0.2887 + 0.50i \hs &0.4783+ 0.4432 i  \\
\end{array}%
\right)\hspace{0.25cm}\,\mbox{for \, IH.}
\end{array}%
\right.
\label{4}
\eea
It is checked that both forms of the leptonic mixing matrix $U_{lep}$ given in
(\ref{4}) are  unitary and consistent with the constraint given in Ref. \cite{Esteban2020}.

As a consequence,  the Jarlskog invariant is given by
\bea
J_{CP}&=& \left\{
\begin{array}{l}
-2.501\times 10^{-2} \hspace{0.3cm}\mbox{for \, NH},    \\
-2.744\times 10^{-2}\hspace{0.25cm}\,\mbox{for \, IH}.
\end{array}%
\right. \hspace{0.675cm}
\eea
From above analysis, we can conclude that the model under consideration can reproduce the recent
experimental values of neutrino mixing angles and Dirac CP violating phase \cite{Esteban2020} in which the
atmospheric angle $(\theta_{23})$ and the reactor angle $(\theta_{13})$ get the best-fit values while the solar angle $(\theta_{12})$
	 and Dirac CP violating phase ($\de $) belong to $3\, \si $ range of the best-fit value for normal hierarchy (NH). For inverted
	  hierarchy (IH), $\theta_{13}$ gets the best-fit value and $\theta_{23}$ together with $\de $ belongs to $1\, \si $ range
	   while $\theta_{12}$ belongs to $3\, \si $ range of the best-fit value taken from Ref. \cite{Esteban2020}.  It is noted that although the model
   results on $t_{12}$ and $\de $ belong to $3\, \si $ range of the best-fit value taken from \cite{Esteban2020}, they
   belong to $2\, \si $ of the best-fit value taken from \cite{Salas2020} and $1\, \si $ of the best-fit value taken from
    SNO and KamLAND collaborations \cite{SNO05, KamLAND08}.
Now we turn to neutrino mass hierarchy.
\vspace{-0.5cm}

\subsubsection{\label{NHsect} Normal spectrum}

Taking into account the best-fit values of the neutrino mass-squared differences for NH given
in Tab. \ref{tab1},  $\De  m^2_{21}=7.42 \times 10^{-5} \mathrm{eV}^2, \, \De  m^2_{31}=2.517\times 10^{-3}\mathrm{eV}^2$,
 we obtain a solution
\bea
\ka _1&=&\fr{\sum m_i}{2}-\fr{c_0}{2}, \hs \ka _2 = \fr{1}{2}\left(\sqrt{\de _{1N}}-\sqrt{\de _{2N}}\right),  \label{k1k2n}\\
c_0&=&\fr{\sum m_i}{3}-132.40 \de _{1N}\sqrt{\de _{1N}}+2.102\times 10^{-5}\left(\sqrt{\de _{1N}}
+\sqrt{\de _{2N}}\right)\sqrt[3]{\de _{N}}\crn
&+&\sqrt{\de _{2N}}\left[132.4\de _{2N}-0.3137-264.90 \left(\sum m_i\right)^2\right]+ \fr{\left(\sqrt{\de _{1N}}  +\sqrt{\de _{2N}}\right)\de _{3N}}{\sqrt[3]{\de _{N}}}, \label{c0n}
\eea
where $\de _N$ and $\de _{iN} \, (i=1\div 4)$ are given in Appendix \ref{expression1}.
Expressions (\ref{m1m2m3}), (\ref{k1k2n}), (\ref{c0n}) and (\ref{D1N}) -- (\ref{D4N}) show that three neutrino masses
$m_{1,2,3}$ depend only on the sum of neutrino masses $\sum_{i=1}^3 m_i$.

At present there are various bounds on $\sum m_{i}$, for instance, for the NH, the upper limit on the sum of neutrino
masses is $\sum m_i  < 0.13\,  \mathrm{eV}$ at $2\,\si $ range \cite{Salas2020}. The dependence of $m_{1,2,3}$ on $\sum m_i $
is plotted in Fig. \ref{m123N} with $\sum m_i \in \left(0.06, 0.1\right)\, \mathrm{eV}$ within $2\, \si $ range of the best-fit value
taken from Ref. \cite{Salas2020} and being well consistent with the strongest bound from
 cosmology \cite{nubound} $\sum m_\nu < 0.078 \, \mathrm{eV}$
and  the upper bounds taken from \cite{Capozzi20} $\sum m_\nu < 0.12\div 0.69 \, \mathrm{eV}$.
\begin{figure}[h]
\begin{center}
\vspace{-1.5 cm}
\hspace{-0.5 cm}\includegraphics[width=0.8\textwidth]{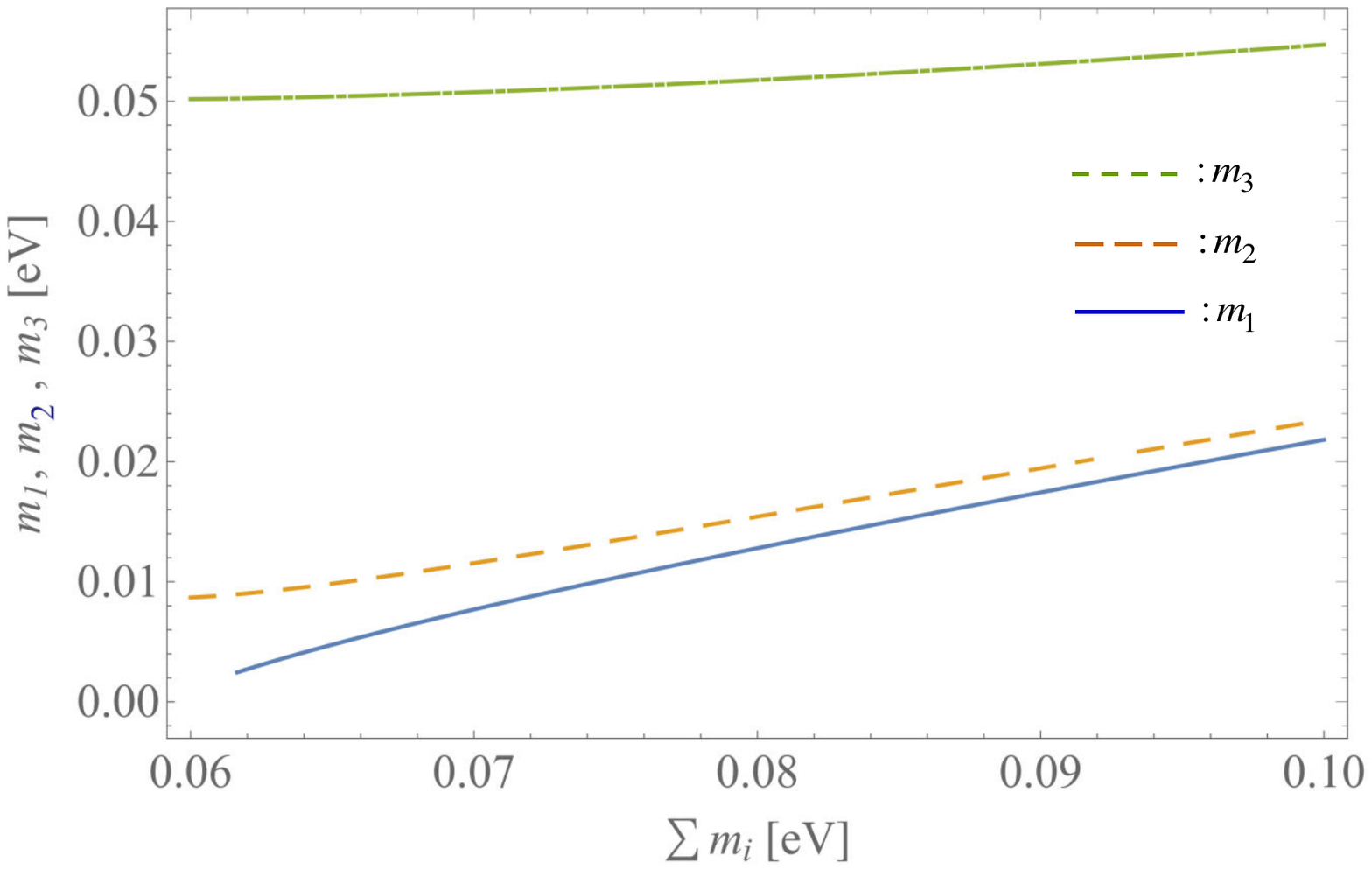}
\end{center}
\vspace{-9.0cm}
\caption{ $m_{1,2,3}$  versus $\sum m_i$ with $\sum m_i \in \left(0.06, 0.1\right)\, \mathrm{eV}$ in the NH.}
\label{m123N}
\end{figure}
In the case $\sum m_{i}=6.5\times 10^{-2}\, \mathrm{eV}$ we get
\bea
&& m_1= 4.76\times 10^{-3}\, \mathrm{eV}, \hs m_2= 9.84\times 10^{-3}\, \mathrm{eV}, \hs m_3=5.04\times 10^{-2}\, \mathrm{eV}. \label{m123N}
\eea
\subsubsection{\label{IHsect} Inverted spectrum}
\begin{figure}[h]
\begin{center}
\vspace{-1.5 cm}
\hspace{-0.5 cm}\includegraphics[width=0.8\textwidth]{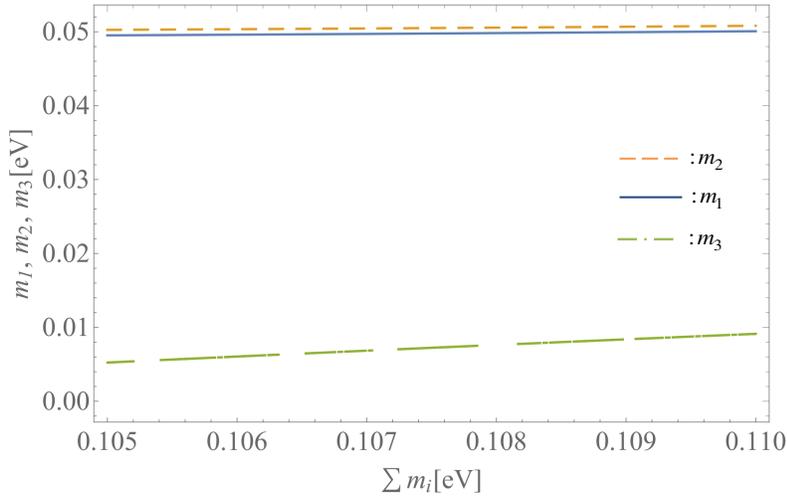}
\end{center}
\vspace{-9.0cm}
\caption{ $m_{1,2,3}$  versus $\sum m_i$ with $\sum m_i \in \left(0.105, 0.11\right)\, \mathrm{eV}$ in the IH.}
\label{m1m2m3I}
\end{figure}

As before, using  the best-fit values of the neutrino mass-squared differences for IH 
 shown in Tab. (\ref{tab1}),  $\De  m^2_{21}=7.42 \times 10^{-5} \mathrm{eV}^2,\,
\De  m^2_{32}=-2.498\times 10^{-3}\mathrm{eV}^2$, we get a solution:
\bea
\ka _1&=&\fr{\sum m_i}{2}-\fr{c_0}{2}, \hs \ka _2 = \fr{1}{2}\left(\sqrt{\de _{1I}}-\sqrt{\de _{2I}}\right),  \label{k1k2i}\\
c_0&=&\fr{\sum m_i}{3}-137.50 \de _{1I}\sqrt{\de _{1I}}+2.183\times 10^{-5}\left(\sqrt{\de _{1I}}
+\sqrt{\de _{2I}}\right)\sqrt[3]{\de _{I}}\crn
&+&\sqrt{\de _{2I}}\left[137.50\de _{2I}+0.3537-275.10 \left(\sum m_i\right)^2\right]+ \fr{\left(\sqrt{\de _{1I}}  +\sqrt{\de _{2I}}\right)\de _{3I}}{\sqrt[3]{\de _{I}}}, \label{c0i}
\eea
where $\de _I$ and $\de _{iI} \, (i=1\div 4)$ are given in Appendix \ref{expression1}.
 Similarly to previous section,  three neutrino
masses $m_{1,2,3}$ just  depend 
 on the sum of neutrino masses $\sum_{i=1}^3 m_i$.
In the IH, the tightest $2\si $ upper limit on the sum of neutrino masses has been reached as follows $\sum m_i  < 0.15 \,  \mathrm{eV}$  \cite{Salas2020}.
Furthermore,  the upper bounds taken from \cite{Capozzi20} is $\sum m_\nu < 0.12\div 0.69\,  \mathrm{eV}$.
The dependence of three active neutrino masses $m_{1,2,3}$ on $\sum m_i $ is plotted in Fig. \ref{m1m2m3I}
with $\sum m_i \in \left(0.1, 0.2\right)\, \mathrm{eV}$ which is well consistent with the recent
constraints given in Refs.\cite{Salas2020,Capozzi20}. In the case $\sum m_{i}=0.1075\, \mathrm{eV}$ we get
\be
 m_1= 4.976\times 10^{-2}\, \mathrm{eV}, \hs m_2= 5.05\times 10^{-2}\, \mathrm{eV}, \hs m_3=7.237\times 10^{-3}\, \mathrm{eV}. \label{m123I}
\ee
\subsubsection{\label{effect}Effective neutrino mass parameters}
\begin{figure}[ht]
\begin{center}
\hspace{-0.4 cm}\includegraphics[width=1.025\textwidth]{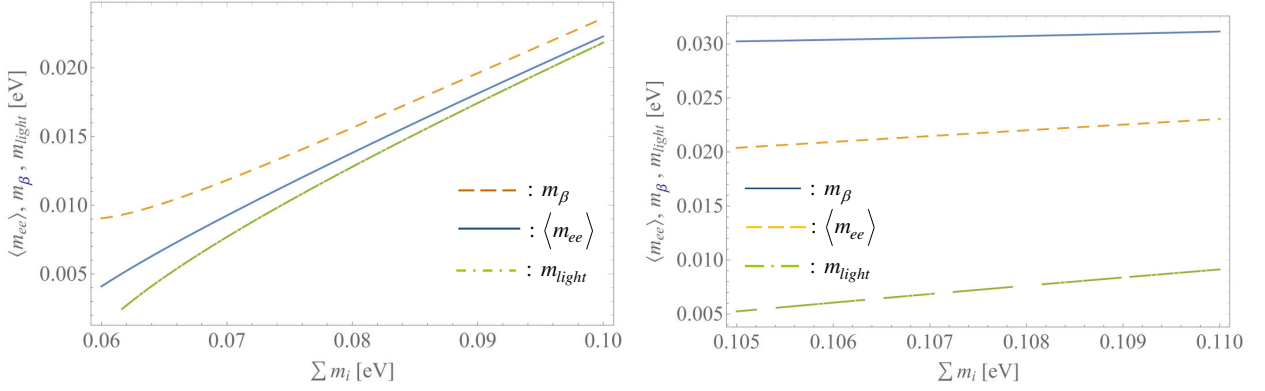}\hspace{-5.5 cm}
\end{center}
\vspace{-7.75cm}
\caption{ $\langle m_{ee} \rangle, m_\beta$ and $m_{light}$ versus $\sum m_i$ with $\sum m_i \in \left(0.06, 0.1\right)\,
\mathrm{eV}$ in the NH (left panel) and $\sum m_i \in \left(0.105, 0.11\right)\, \mathrm{eV}$ in the IH (right panel).}
\label{meembNI}
\end{figure}

Now, we deal with an  effective neutrino mass. Expressions (\ref{m1m2m3}), (\ref{k1k2n}), (\ref{c0n}) and (\ref{D1N}) -- (\ref{D4N}) (for NH)
and (\ref{m1m2m3}), (\ref{k1k2i}), (\ref{c0i}) and (\ref{D1I}) -- (\ref{D4I}) (for IH) show that, with the best-fit
 values of the neutrino mass-squared differences, the effective neutrino mass parameters $\langle m_{ee}\rangle$
 and $m_\beta$ depend on the sum of neutrino masses $\sum_{i=1}^3 m_i$ and two mixing
 angles $\theta_{23},\, \theta_{13}$. In the NH, $m_1< m_2<m_3$, hence $m_1\equiv m_{light}$ is
  the lightest neutrino mass while in the IH, $m_3< m_1< m_2$, therefore $m_3\equiv m_{light}$ is the
  lightest neutrino mass. If we fix $\theta_{23}$ and $\theta_{13}$ at their best-fit values taken
    in Table \ref{tab1}, the effective neutrino masses
    $\langle m_{ee}\rangle, m_{\beta}$ and $m_{light}$ as functions of $\sum m_i$ has been plotted in Fig.\ref{meembNI}.

\begin{figure}[h]
\begin{center}
\vspace{-0.15 cm}
\hspace{-5.5 cm}
\includegraphics[width=0.65\textwidth]{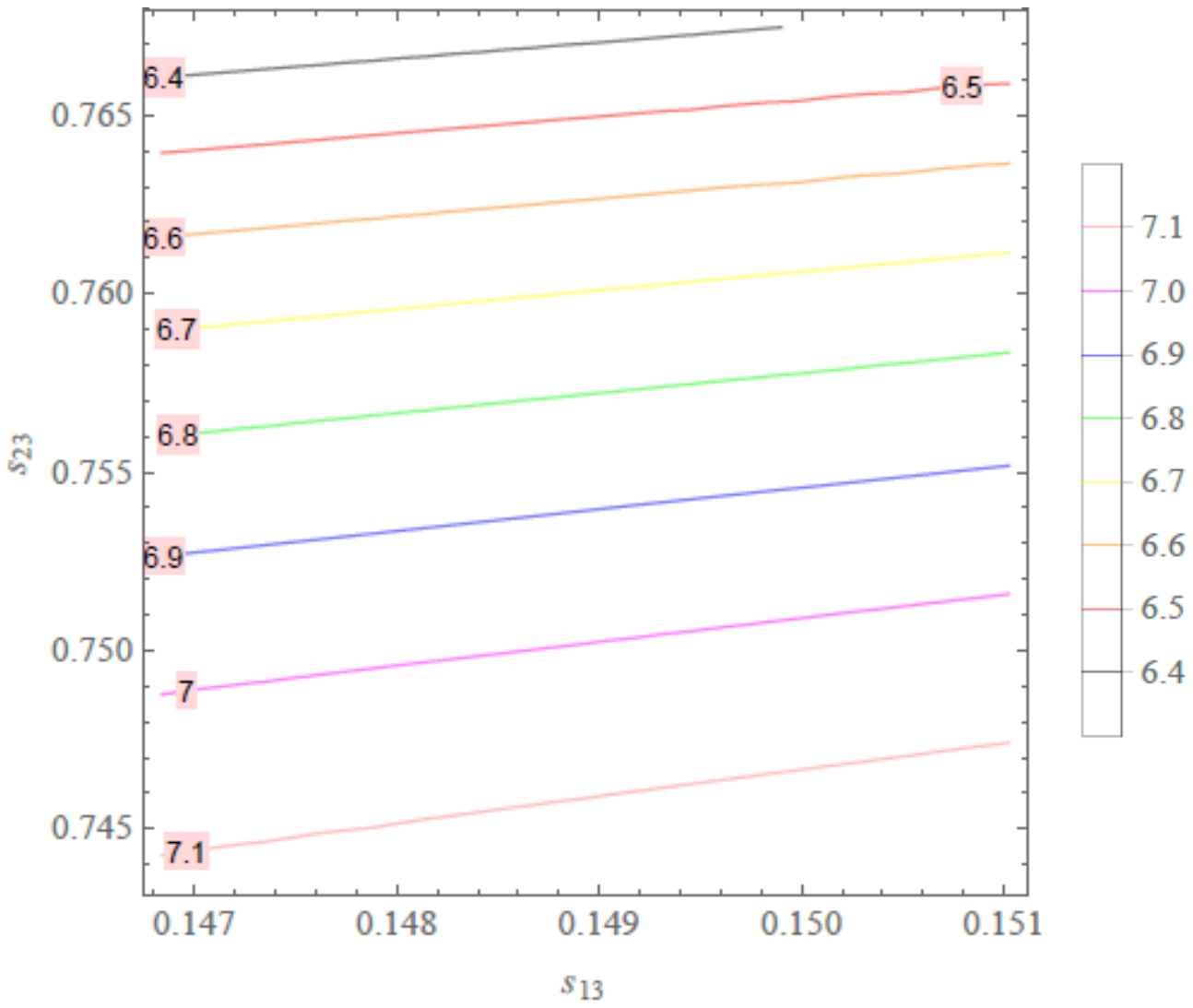}\hspace{-4.0 cm}
\includegraphics[width=0.65 \textwidth]{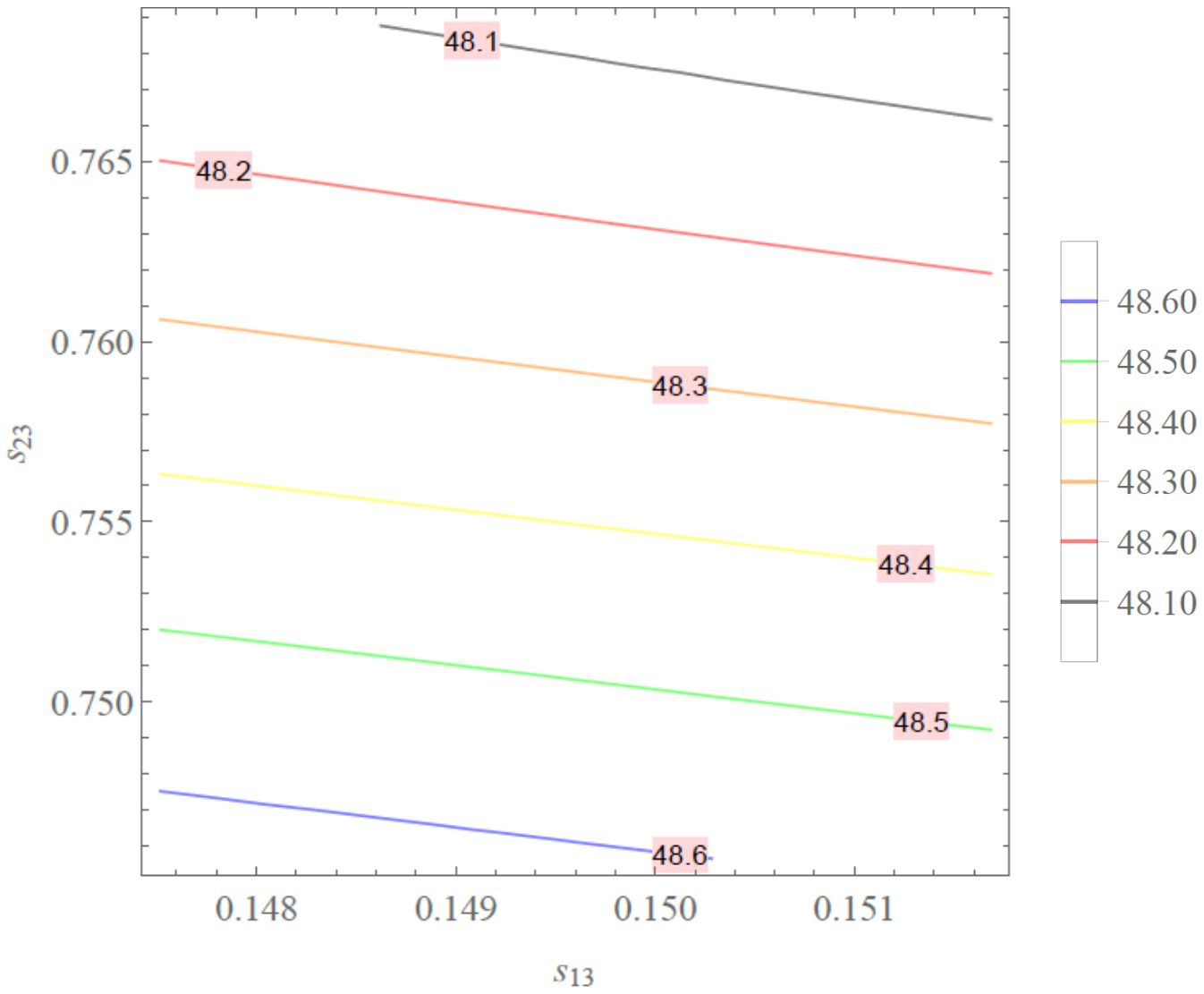}\hspace*{-6.5 cm}
\end{center}
\vspace{-7.75 cm}
\caption[The contour plot $\langle m_{ee}\rangle \, (\mathrm{meV})$ as a function of the $s_{13}$ and $s_{23}$ in the $1\si $ range
 of the best-fit value given \crb{in Table \ref{tab1},} i.e., $s_{13}\in (0.1468,\, 0.1510)$ and $s_{23}\in (0.7436,\, 0.7675)$
 for NH (left panel), and $s_{13}\in (0.1475,\, 0.1517)$ and $s_{23}\in (0.7457,\, 0.7688)$ for IH (left panel).]{The contour
  plot $\langle m_{ee}\rangle \, (\mathrm{meV})$ as a function of the $s_{13}$ and $s_{23}$ in the $1\si $ range of the best-fit values:
   $s_{13}\in (0.1468,\, 0.1510)$ and $s_{23}\in (0.7436,\, 0.7675)$ for NH (left panel),
  and $s_{13}\in (0.1475,\, 0.1517)$ and $s_{23}\in (0.7457,\, 0.7688)$ for IH (left panel).}
\label{mee}
\end{figure}

\begin{figure}[h]
\begin{center}
\vspace{-0.25 cm}
\hspace{-5.0 cm}
\includegraphics[width=0.65\textwidth]{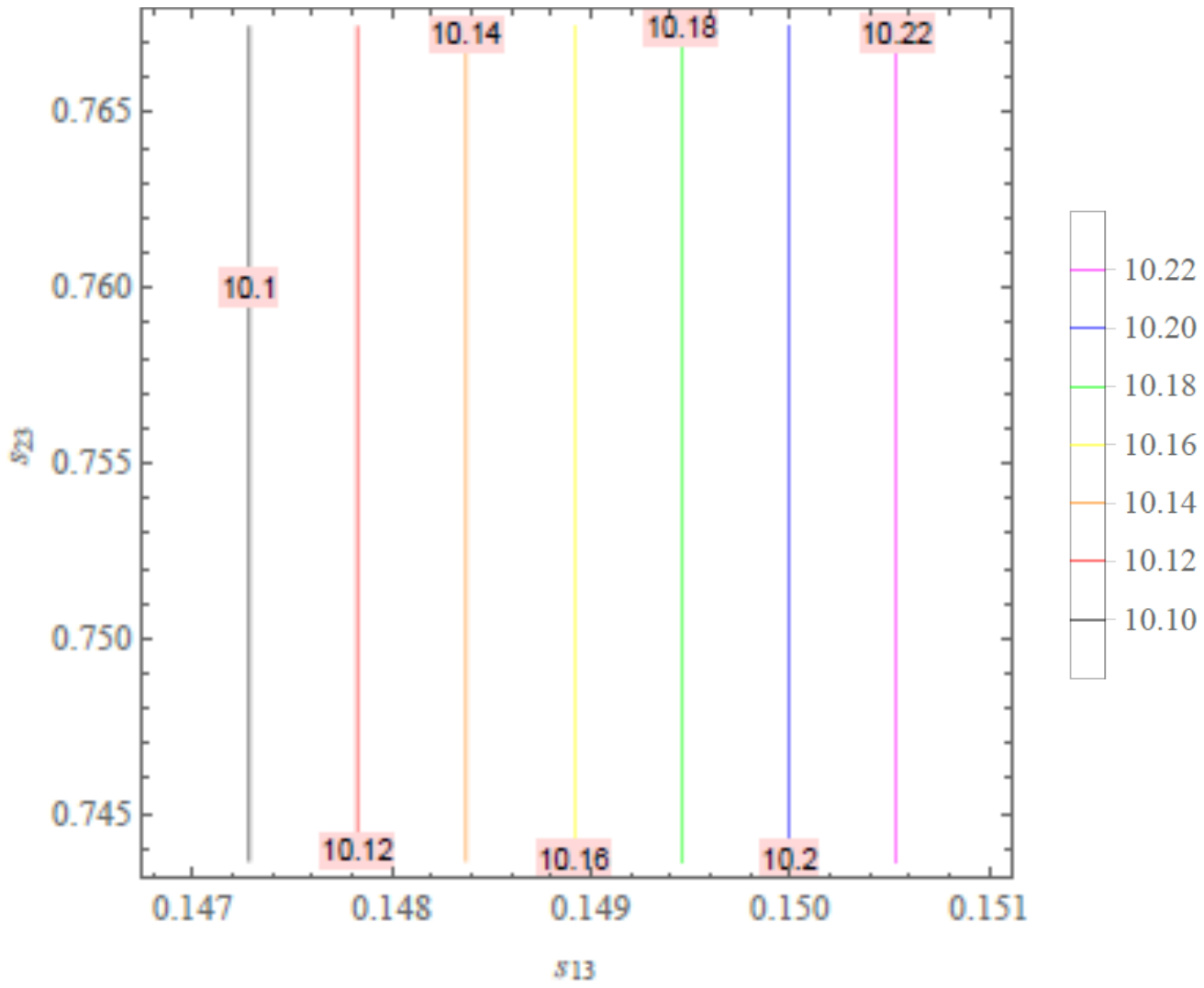}\hspace{-3.75 cm}
\includegraphics[width=0.65\textwidth]{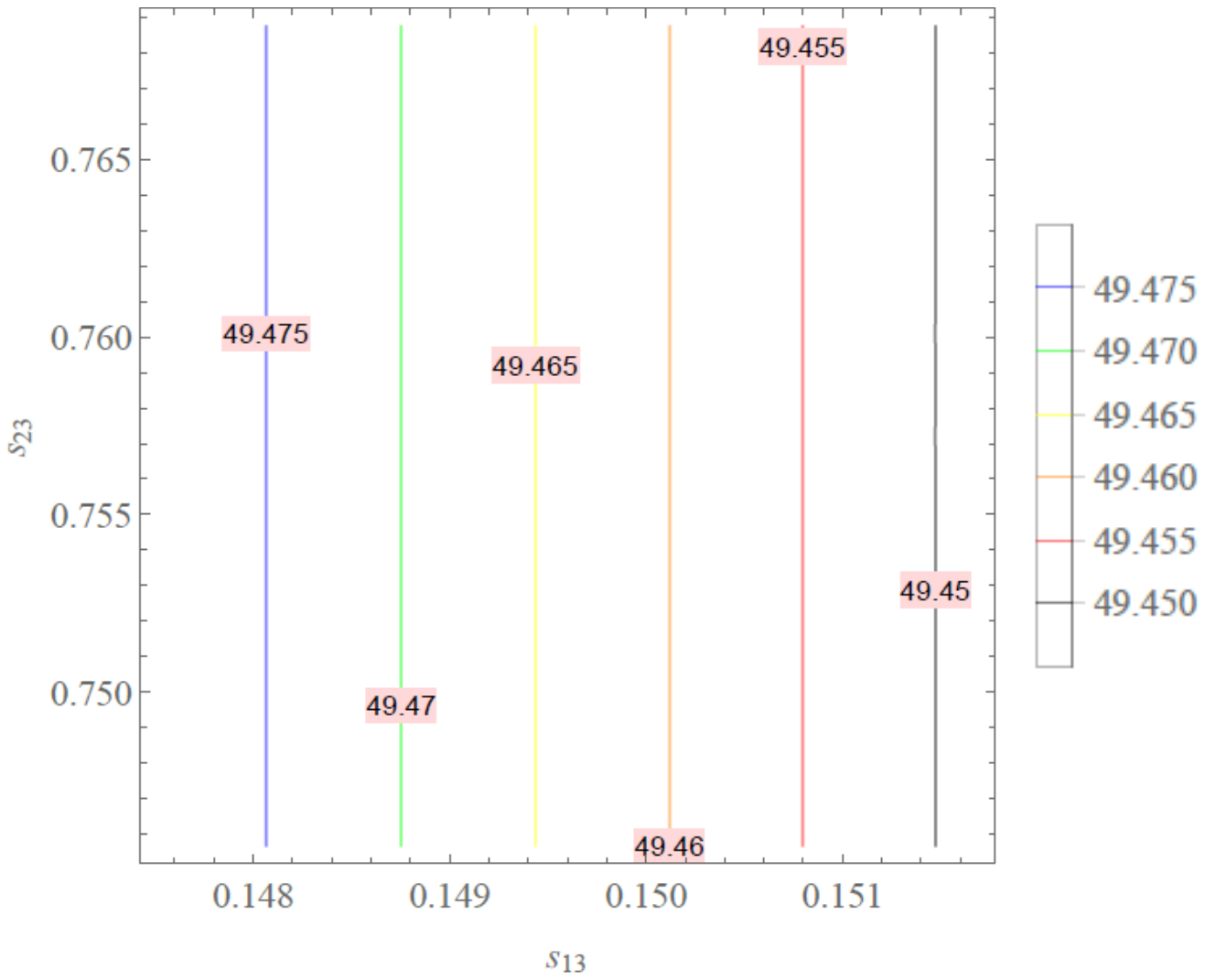}\hspace*{-5.5 cm}
\end{center}
\vspace{-8.25 cm}
\caption[The contour plot $m_\beta \, (\mathrm{meV})$ as a function of the $s_{13}$ and $s_{23}$ in the $1\si $ range of the best-fit value given in Table \ref{tab1}:
 $s_{13}\in (0.1468,\, 0.1510)$ and $s_{23}\in (0.7436,\, 0.7675)$ for NH (left panel), and $s_{13}\in (0.1475,\, 0.1517)$
and $s_{23}\in (0.7457,\, 0.7688)$ for IH (left panel).]{The contour plot $m_\beta \, (\mathrm{meV})$ as a function of the $s_{13}$ and $s_{23}$
 in the $1\si $ range of the best-fit values:
  $s_{13}\in (0.1468,\, 0.1510)$
  and $s_{23}\in (0.7436,\, 0.7675)$ for NH (left panel), and $s_{13}\in (0.1475,\, 0.1517)$ and $s_{23}\in (0.7457,\, 0.7688)$ for IH (left panel).}
\label{mb}
\end{figure}

In order to see the dependence  of $\langle m_{ee} \rangle$ and $m_\beta$ on $\theta_{23}$ and $\theta_{13}$
we can fix the value for $\sum m_i$ in its constraint range \cite{Salas2020,Capozzi20}, for instance, $\sum m_i=0.065\, \mathrm{eV}$
for NH and $\sum m_i=0.1075\, \mathrm{eV}$ for IH. Consequently,  we can contour plot $\langle m_{ee} \rangle$ and $m_\beta$ as
 functions of ($\theta_{23},\,\theta_{13}$) as shown in Figs. \ref{mee} and \ref{mb}, respectively.
\newline
These figures show that  at $1\, \si $ range of the best-fit value taken from Ref. \cite{Esteban2020} of $s_{23}$ and $s_{13}$,
 the model predicts the range of the effective neutrino mass parameters as follows
\bea
\langle m_{ee}\rangle \in\left\{
\begin{array}{l}
(6.40,\,\, 7.10) \,\, \mbox{meV}\ \ \ \ \  \mbox{for \ \  NH,} \\
(48.04,\, 48.64)\,\, \mbox{meV}\ \  \mbox{for \ \ \  IH,}%
\end{array}%
\right.   \label{meeranges}
\eea
and \bea
m_{\beta}\in\left\{
\begin{array}{l}
(10.10,\,\, 10.22) \,\, \mbox{meV}\ \ \ \ \ \  \mbox{for \ \ \ \ NH,} \\
(49.45,\,49.48)\,\, \mbox{meV}\ \ \ \  \mbox{for \ \ \ \ IH.}%
\end{array}%
\right.   \label{mbranges}
\eea
In the case $s_{23}$ and $s_{13}$ take their best-fit values \cite{Esteban2020}, $s_{23}=0.757,\, s_{13}=0.149$
for NH and for $s_{23}=0.7583,\, s_{13}=0.1496$IH,
ones get:
\bea
\langle m_{ee}\rangle =\left\{
\begin{array}{l}
6.81 \,\, \mbox{meV}\ \ \ \ \  \mbox{for \ \ \ NH,} \\
48.48 \,\, \mbox{meV}\ \ \ \  \mbox{for \ \ \  IH,}%
\end{array}%
\right.   \label{meetvalues}
\eea
and \bea
m_{\beta}=\left\{
\begin{array}{l}
10.20 \,\, \mbox{meV}\ \ \ \ \   \mbox{for \ \ \  NH,} \\
49.46\,\, \mbox{meV}\ \ \ \  \mbox{for \ \ \  IH.}%
\end{array}%
\right.   \label{mbetvalues}
\eea
The derived effective neutrino mass parameters in Eqs.(\ref{meetvalues}) and (\ref{mbetvalues}) satisfy all the upper bounds arising from recent $0\nu \beta \beta $ decay experiments taken from
KamLAND-Zen \cite{KamLAND16} $\langle m_{ee} \rangle <61\div 165 \,\mathrm{meV}$, GERDA \cite{GERDA19}
 $\langle m_{ee} \rangle < 104 \div 228 \,\mathrm{meV}$ and CUORE \cite{CUORE20} $\langle m_{ee} \rangle < 75 \div 350 \,\mathrm{meV}$.

\section{\label{conclusion} Conclusions}
We have suggested a multiscalar and nonrenormalizable $U(1)_{B-L}$ extension of the SM with
$S_4$ symmetry
 which successfully explains
 the recent observed neutrino oscillation data. The tiny neutrino
 and the neutrino masses hierarchy are generated via
  the type-I seesaw mechanism. The model reproduces the recent
 experimental data of neutrino mixing angles and Dirac CP violating phase in which the atmospheric angle $(\theta_{23})$
  and the reactor angle $(\theta_{13})$ get the best-fit values while the solar angle $(\theta_{12})$ and Dirac CP violating phase ($\de $) belong to $3\, \si $ range of the best-fit value for NH. For IH, $\theta_{13}$ gets the best-fit value and $\theta_{23}$ together with $\de $ belongs to $1\, \si $ range while $\theta_{12}$ belongs to $3\, \si $ range of the best-fit value.
The effective neutrino masses are predicted to be $\langle m_{ee}\rangle =6.81\, \mathrm{meV}$ for NH and $\langle m_{ee}\rangle= 48.48\,\mathrm{meV}$ for  IH while $m_{\beta} =10.20\, \mathrm{meV}$ for NH and $m_{\beta}= 49.46\,\mathrm{meV}$ for  IH which are very  well consistent with the most recent experimental data.
\section*{Acknowledgments}
This research is funded by Vietnam National Foundation for Science and Technology Development (NAFOSTED) under grant number 103.01-2017.341.
\appendix
\section{\label{expression1} The explicit expression of $\de _{N(I)}$ and $\de _{iN(I)} \, (i=1\div 4)$}
The parameters $\de _{N(I)}$ and $\de _{iN(I)} \, (i=1\div 4)$ in Eqs. (\ref{k1k2n}), (\ref{c0n}),  (\ref{k1k2i}), (\ref{c0i})
have the explicit expressions as follows:
\bea
\de _{1N}&=&7.895\times 10^{-4}+5.291\times 10^{-8}\sqrt[3]{\de _{N}}+\fr{2}{3}\left(\sum m_i\right)^2 \crn
&-&\fr{26.99-4979\left(\sum m_i\right)^2-2.10\times 10^6 \left(\sum m_i\right)^4}{\sqrt[3]{\de _{N}}}, \label{D1N} \\
\de _{2N}&=&1.5795\times 10^{-3}+5.291\times 10^{-8}\sqrt[3]{\de _{N}}+\fr{4}{3}\left(\sum m_i\right)^2\crn
&+&\fr{26.99-4979\left(\sum m_i\right)^2-2.10\times 10^6 \left(\sum m_i\right)^4}{\sqrt[3]{\de _{N}}}+\fr{5.034\times 10^{-3} \sum m_i}{\sqrt{\de _{1N}}}, \\
\de _{3N}&=&-1.0722\times 10^4 + 1.976\times 10^6 \left(\sum m_i\right)^2 + 8.343\times 10^8 \left(\sum m_i\right)^4,\\
\de _{N}&=&-1.308\times 10^{13}+9.639\times 10^{15}\left(\sum m_i\right)^2 -8.882\times 10^{17}\left(\sum m_i\right)^4 \crn
&-& 2.50\times 10^{20}\left(\sum m_i\right)^6 -6.539\times 10^{12}\sqrt{\de _{4N}},\\
\de _{4N}&=&7.101-7.61 \times 10^3\left(\sum m_i\right)^2+2.308\times 10^6\left(\sum m_i\right)^4-6.65\times 10^{10}\left(\sum m_i\right)^{8}. \label{D4N}\\
\de _{1I}&=&-8.574\times 10^{-4}+5.291\times 10^{-8}\sqrt[3]{\de _{I}}+\fr{2}{3}\left(\sum m_i\right)^2 \crn
&-&\fr{24.28+5401\left(\sum m_i\right)^2-2.10\times 10^6 \left(\sum m_i\right)^4}{\sqrt[3]{\de _{I}}}, \label{D1I} \\
\de _{2I}&=&-1.715\times 10^{-3}-5.291\times 10^{-8}\sqrt[3]{\de _{I}}+\fr{4}{3}\left(\sum m_i\right)^2\crn
&+&\fr{24.28+5401\left(\sum m_i\right)^2-2.10\times 10^6 \left(\sum m_i\right)^4}{\sqrt[3]{\de _{I}}}-\fr{4.848\times 10^{-3} \sum m_i}{\sqrt{\de _{1I}}}, \eea
\bea
\de _{3I}&=&-1.002\times 10^4 -2.228\times 10^6 \left(\sum m_i\right)^2 + 8.664\times 10^8 \left(\sum m_i\right)^4,\\
\de _{I}&=&1.328\times 10^{13}+8.673\times 10^{15}\left(\sum m_i\right)^2 +9.646\times 10^{17}\left(\sum m_i\right)^4 \crn
&-&2.50\times 10^{20}\left(\sum m_i\right)^6 -6.297\times 10^{12}\sqrt{\de _{4I}},\\
\de _{4I}&=&6.886+7.436\times 10^3\left(\sum m_i\right)^2+2.273\times 10^6\left(\sum m_i\right)^4-6.25\times 10^{10}\left(\sum m_i\right)^{8}.\hs\,\,\label{D4I}
\eea

\end{document}